\documentclass[]{raa}    % referee version: for submission

\usepackage{graphicx,times}             %for PS/EPS graphics inclusion, new
\usepackage{natbib}
\usepackage{amssymb,amsmath}
\bibpunct{(}{)}{;}{a}{}{,}

\usepackage[pagebackref=true]{hyperref}

\begin{document}

  \title{$Herschel$ investigation of cores and filamentary structures in the Perseus molecular cloud}

%   \subtitle{I. Place Your Subtitle Here}

   \volnopage{Vol.0 (20xx) No.0, 000--000}      %%preserved for Editor. DOn't remove!
   \setcounter{page}{1}          %%starting page, preserved for Editor. DOn't remove!

   \author{Chang Zhang %(周爱英) %% Put your Chinese name in "( )" if you like. Note to open line 11 "\usepackage[UTF8]{ctex}"
      \inst{1,2}
   \and Guo-Yin Zhang 
      \inst{1}
   \and Jin-Zeng Li
      \inst{1}
   \and Xue-Mei Li
      \inst{3}   
   }
%% Here is an example of three authors come from different institutes.
%% For single author or all the authors from an institute, use "\inst{}" only

   \institute{National Astronomical Observatories, Chinese Academy of Sciences, Beijing 100101, PR China; {\textit zhangc@nao.cas.cn; zgyin@nao.cas.cn;  ljz@nao.cas.cn}\\
%% Please give the E-mail address of the author, to whom future correspondence and
%% offprint requests will be sent.
        \and
        University of Chinese Academy of Sciences, Beijing 100049, PR China\\
        \and
        College of Physics, Guizhou University, Guiyang 550025, PR China\\     
\vs\no
   {\small Received 20xx month day; accepted 20xx month day}}

\abstract{Cores and filamentary structures are the prime birthplaces of stars, 
and play key roles in the process of star formation.
Latest advances in the methods of multi-scale source and filament 
extraction, 
and in making high-resolution column density map from $Herschel$ multi-wavelength 
observations enable us to detect the filamentary network structures in highly 
complex molecular cloud environments. The statistics for physical parameters 
shows that core mass strongly correlates with core dust temperature, 
and $M/L$ strongly correlates with $M/T$, which is in line with the 
prediction of the blackbody radiation, and can be used to trace 
evolutionary sequence from unbound starless cores to robust 
prestellar cores. 
Crest column densities of the filamentary structures are clearly related 
with mass per unit length ($M_{\rm line}$), but are 
uncorrelated by three orders ranging from $\sim 10^{20}$ to $\sim 10^{22}$ $ \rm cm^{-2}$ with widths.
Full width at half maximum (FWHM) have a median value of 0.15 pc, 
which is consistent with the 0.1 pc typical inner width of the 
filamentary structures reported by previous research.
We find $\sim $70\% of robust prestellar cores (135/199) 
embedded in supercritical filaments with $M_{\rm line}>16~M_{\odot}/{\rm pc}$, 
which implies that the gravitationally bound cores come from fragmentation of 
supercritical filaments.
And on the basis of observational evidences that probability distribution function (PDF) with
power-law distribution in the Perseus south is flatter 
than north, YSO number is significantly less than that in the north, 
and dust temperature difference.
We infer that south region is more gravitationally bound 
than north region.
}
   \authorrunning{C. Zhang, et al. }            %author_head in even pages
   \titlerunning{Herschel investigation of cores and filamentary structures in the Perseus molecular cloud }  % title_head in odd pages

   \maketitle
%% The author head (on even pages) and the title head (on odd pages) will be
%% automatically extracted from \author{} and \title{}. Whenever the title is too long,
%% you will be asked to supply a shorter one by inserting either \authorrunning{} or
%% \titlerunning{} before \maketitle. Anyway, you can specify your own heads.
%%
%%
%% Note: In the following text body of your manuscript, please note several differences from
%%       other major journals:
%% (1) \subsection{Please Capitalize the First Letter of Each Notional Word in Subsection Title}
%% (2) Please Capitalize the First Letter of Each Notional Word in all tables' captions

%
%________________________________________________ sections below
%
\section{Introduction}           %% first-level sections will be auto-capitalized
\label{sect:intro}

Molecular clouds (MCs) are dense regions of the cold interstellar medium (ISM), 
mainly composed of gas and dust, which are the cradles of stars 
\citep[e.g.][]{BerginandTafalla+2007, Andre+etal+2014, Zhang+etal+2018}.
Molecular clouds are hierarchical in structure \citep{Men+etal+2010,Pokhrel+etal+2018,Men+2021a}.
The substructure of MCs is a complex pattern consisting of filaments, cores, large scale background, local fluctuation, etc.
The detection for core and filamentary structure  and statistics for their 
physical parameters help us understand the initial condition for star formation.

$Herschel's$ far-infrared (FIR) observations of thermal radiation 
from dust provide an unprecedented opportunity to study the 
substructure of molecular clouds and thereby demystify star 
formation \citep[e.g.][]{Konyves+etal+2015,Arzoumanian+etal+2019,Zhang+etal+2020}.
Space telescopes avoid the absorption, distortion and contamination 
of light by Earth's atmosphere. The detected cores with $Herschel$ can be 
an order of magnitude more in number than ground-based 
telescopes \citep[e.g.][]{Konyves+etal+2015,Zhang+etal+2015}.
Observations with $Herschel$ show that filamentary structures at temperatures 
around 10 to 20 K are indeed ubiquitous in the cold interstellar medium.
Results from nearby ($<$ 500 pc) star-forming molecular clouds survey show that 
more than 75\% of prestellar cores are found in supercritical filamentary structures
(Linear density $M_{\rm line}>16\ M_{\odot }\ \rm pc^{-1}$) \citep{Konyves+etal+2015}
and the typical inner width of the filamentary structure is 0.1 pc which is independent 
of the column density \citep{Arzoumanian+etal+2011,Arzoumanian+etal+2019}.
In supercritical filaments observations have revealed quasi-periodic chains of 
dense cores with spacing of 0.15 parsec comparable to the filament inner width 
by \citet{Zhang+etal+2020}. This implies dense filaments will fragment into gravitationally 
bound cores, most of which can evolve into stars.
The detailed fragmentation manner of the filaments may be controlled 
by $M_{\rm line}$, geometrical bending, continuous accretion of gas, 
and magnetic fields \citep[e.g.][]{Zhang+etal+2020}.

\begin{figure*}
   \centering
   \includegraphics[width=\textwidth]{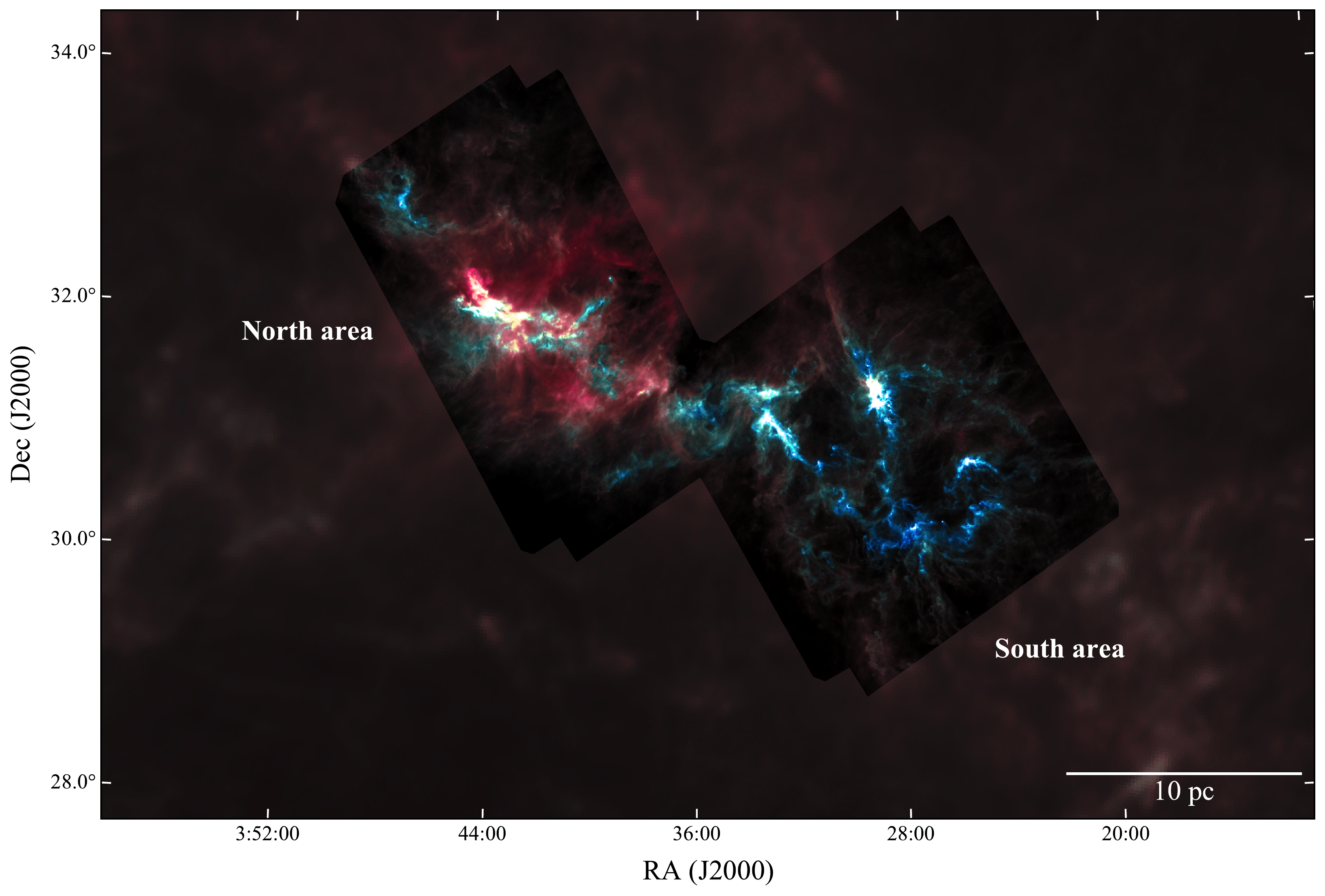}
   \caption{Red-Green-Blue (RGB) composites image showing the 
   250 $\mu$m (blue), 350 $\mu$m (green), and 500 $\mu$m (red) SPIRE fluxes 
   for the Perseus molecular cloud derived from 
   $Herschel$ and $Planck$ observations. 
   In areas outside the $Herschel$ coverage, dust models 
   were applied to predict the corresponding SPIRE fluxes using $Planck$/$IRAS$ data. 
   $Planck$ data have a lower resolution than $Herschel's$, allowing us to identify
   their spatial boundaries. 
   }
   \label{threecolor}
    \end{figure*}

The distance of Perseus MC is $\sim 294$ pc \citep{Zucker+etal+2019}. 
The overall structure can be divided into north and south parts (See Figure \ref{threecolor}). 
The Perseus MC contain several star-forming dust condensations such as B1, B5, 
IC348, NGC1333, L1455 and L1448 \citep{Zari+2016}. About 300 young stellar 
objects (YSOs) have been identified in Perseus MC \citep{Mercimek+etal+2017}. 
Different from convolving all maps to $Herschel's$ lowest low resolution 
of ${36.3}''$ and then fitting spectral energy distribution (SED) to get a column density map to study Perseus MC, 
or single-dish low-resolution molecular line mapping \citep[e.g.][]{Ridge+2006, 
Sadavoy+2014}, that resolution is not high enough to get the full sample
of cores and filaments, we adopted high-resolution column density images 
that derived using an improved difference term algorithm that uses all unresolved or 
slightly resolved structures for enhanced contrast \citep{Men+2021a}.
And latest advances in the methods of multi-scale source and filament
extraction: $getsf$ \citep{Men+2021a}, enable us to detect the sources and filamentary 
network structures in highly complex MC environments and 
perform statistical analysis on physical parameters such as, source luminosity, 
mass, size, and filament width, linear density, curvature etc, as well as 
exploring the correlation of various parameters.

The outline of the present paper is as follows. 
In Sect. \ref{sect:DataandObs}, we describe the $Herschel$ submillimeter dust 
emission data FCRAO 14m molecular line observations of the Perseus MC and give a 
brief overview for $getsf$ . Results are presented in Sect.~\ref{Sect:Resu}. 
In Sect.~\ref{sec:discussion}, 
we discuss different evolution stages of the north and south of the 
Perseus MC, core evolution, as well as characteristic physical parameters of 
filamentary structure. We summarize our conclusions in Sect.~\ref{sect:conclusion}.

%% Authors can give a citation as 'Michel et al. 1992'.
%% You may also use \cite, \citep and \citet for citation, and use Table~1 or figure*~1
%% and so forth. Using \ref and \label for cross-references of Tables/figure*s
%% is a good way in adjusting/adding/removing text, tables or figure*s.

\section{Data reduction and observations}
\label{sect:DataandObs}
\subsection{$Herschel$ Archive data}
  \begin{figure*}
   \centering
   \includegraphics[width=\textwidth]{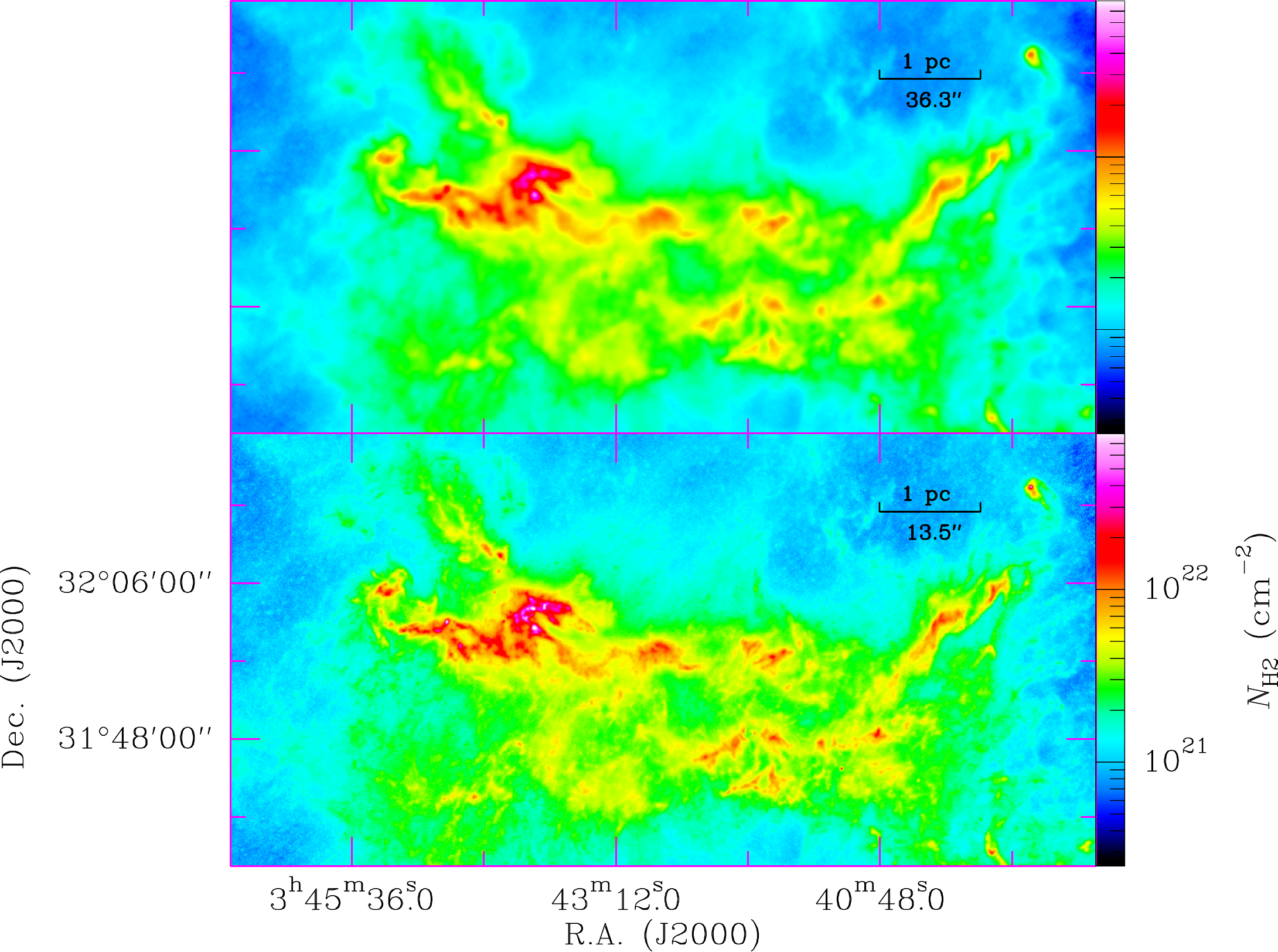}
   \caption{${13.5}''$ and ${36.3}''$ resolution comparison.
   The top panel is a column density map with a resolution of ${36.3}''$.
   The bottom panel is a high-resolution column density image (${13.5}''$) which is derived using an improved difference term algorithm. This algorithm uses all unresolved or slightly resolved structures for enhanced contrast.}
   \label{resocomp}
    \end{figure*}

The $Herschel$ imaging observations of the Perseus molecular cloud 
include PACS 70 $\mu$m and 160 $\mu $m \citep{Poglitsch+etal+2010}
and SPIRE 250, 350, and 500 $\mu $m \citep{Griffin+etal+2010}. 
The beam sizes of the PACS data at 70 and 160 $\mu $m are 8.4 and ${13.5}''$, 
respectively. The beam sizes of the SPIRE data at 250, 350, and 500 $\mu $m 
are 18.2, 24.9, and ${36.3}''$, respectively. 
The SPIRE PACS Parallel Mode with a scanning speed of of ${60}''$/s
is adopted to simultaneously observe this this large area of sky 
in orthogonal mapping directions for the above five bands.
We downloaded $Herschel$ maps from ESA $Herschel$ Science Archive
\footnote{http://archives.esac.esa.int/hsa/whsa/}.
The observed identifiers are 1342190326, 1342190327, 1342214504,1342214505.
Different from $Herschel's$ high-resolution observations,  
$Planck$ and $IRAS$ sacrifice resolution for all-sky observations.
$Planck$ and $IRAS$ data are set as a reference benchmark, then use the 
blackbody radiation model to derive the various bands of $Herschel$, 
and compare with the $Herschel$ data to obtain the Zero-level offsets .
They are 40, 242.9, 96.8, 36.7 and 10.6 MJy/sr at $Herschel$ 160, 250, 350, 
and 500 $\mu $m, respectively in the south region, and 
-3.2,  37.5, -0.6,-0.9 and -0.1 in the north region.
Pixel-by-pixel SED fitting to the $Herschel$ 160–500 $\mu $m data 
with a modified blackbody function was used to create a 
high-resolution (${13.5}''$) $\rm H_{2}$ column density map 
with the method described in the $getsf$ paper \citep{Men+2021a}.
Our column density map has a resolution that is triple as high as the 36.3$''$-resolution 
map commonly used in previous studies, and can help us see more details of the MC structure. There is a comparison in Fig~\ref{resocomp}.

\subsection{Sources and filamentary structure detection algorithm: $getsf$}
We use $getsf$ to extract sources and filaments by separating their structural 
components in multi-wavelength astronomical images \citep{Men+2021a}.
The algorithm has been validated using a set of benchmark images \citep{Men+2021b}.
Here is a brief introduction of the data processing steps of $getsf$. 
(1), we need to cut the multi-band dust continuum images observed by $Herschel$ 
into images with same pixel size, number of pixels and coordinate system, 
with $getsf$'s built-in script: $prepobs$.
(2), then we fit the SED pixel by pixel with $getsf$'s built-in script: $hires$, 
to obtain a set of column density maps and temperature maps with resolution 
at Herschel each observed band.
(3), next, $getsf$ uses spatial separation techniques to separate the source and 
filamentary structures from each other and remove their large-scale background.
There are also local fluctuations and residual noise in the source and filamentary 
structure images, $getsf$ will use the flattening technique to remove them. 
After the above process of background removal and flattening, source and filamentary 
structures have been cleaned on the single scale images. 
(4), next, $getsf$ will combine single scales together at each wavelength observed by $Herschel$. 
(5), then $getsf$ will detect location of the sources and the skeleton of the 
filamentary structures in the combined images. (6), Finally, $getsf$ will 
measure properties and create catalog of detected sources and filamentary 
structures with its built-in script: $smeasure$ and $fmeasure$.
$getsf$ is a almost fully automatic algorithm, and the parameters 
in the configuration file are the optimal choices after extensive testing. 
The only user input required is the maximum size of the  the structure 
that the user wants to extract.

\section{Results}
\label{Sect:Resu}
\subsection{Physical environment of Perseus}\label{3.1}
A high-resolution column density map (${13.5}''$) can be derived, 
using hires in $getsf$. This map is helpful for better 
detection and deblending of dense structures. 
The zero offset level of the north and south regions is very different.
Therefore the column density distribution maps of the north and south regions 
are made respectively. The total mass is $\sim 1.2\times 10^{4}$ $M_{\odot }$ 
obtained by adding up the value of each pixel in the column density map 
and then multiplying by the mass and weight of the hydrogen molecule.
The masses in north and south regions are $\sim 5\times 10^{3}$ $M_{\odot }$ 
and  $\sim 7\times 10^{3}$ $M_{\odot }$ respectively.

  \begin{figure*}
   \centering
   \includegraphics[width=0.48\textwidth]{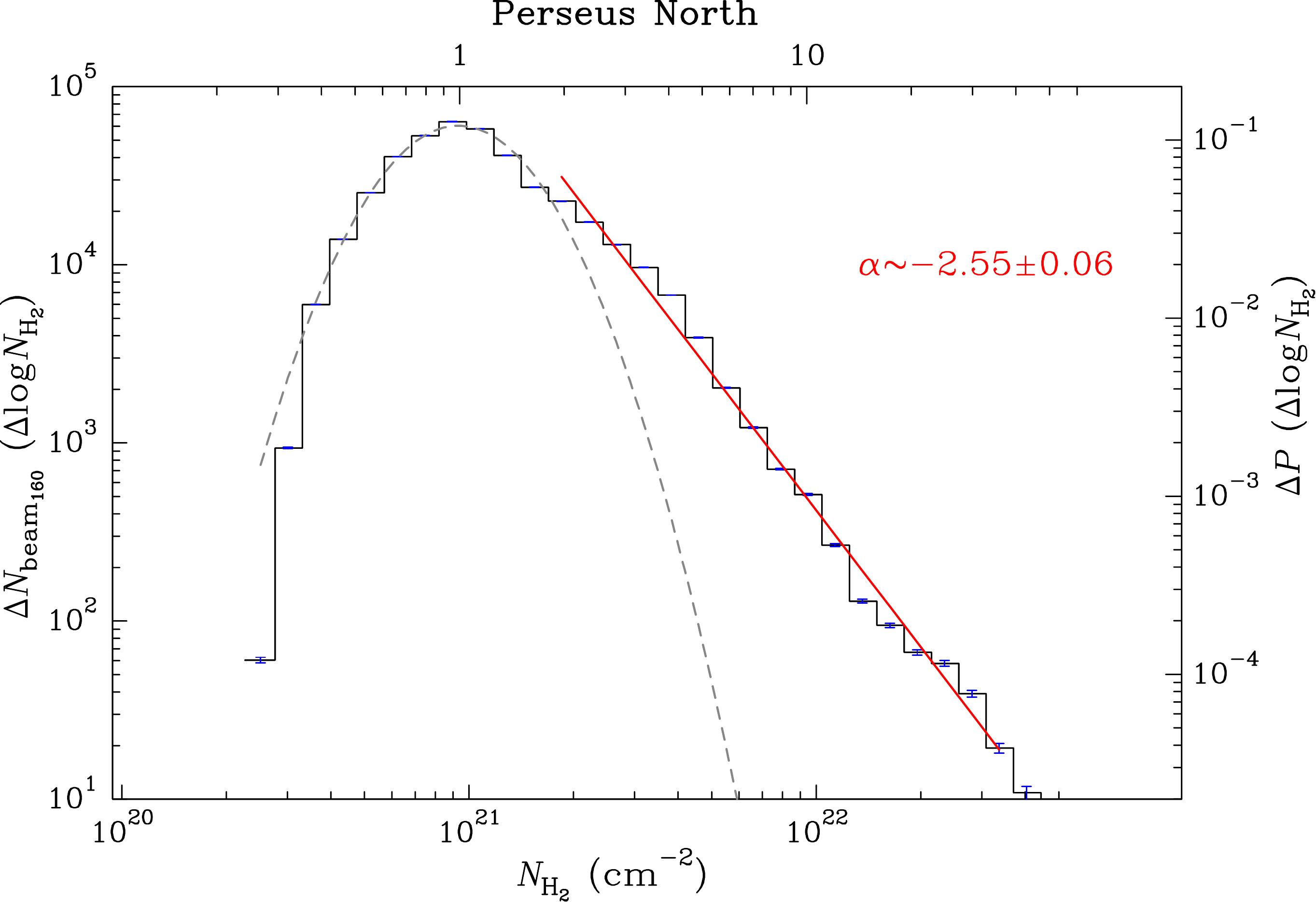}
   \includegraphics[width=0.48\textwidth]{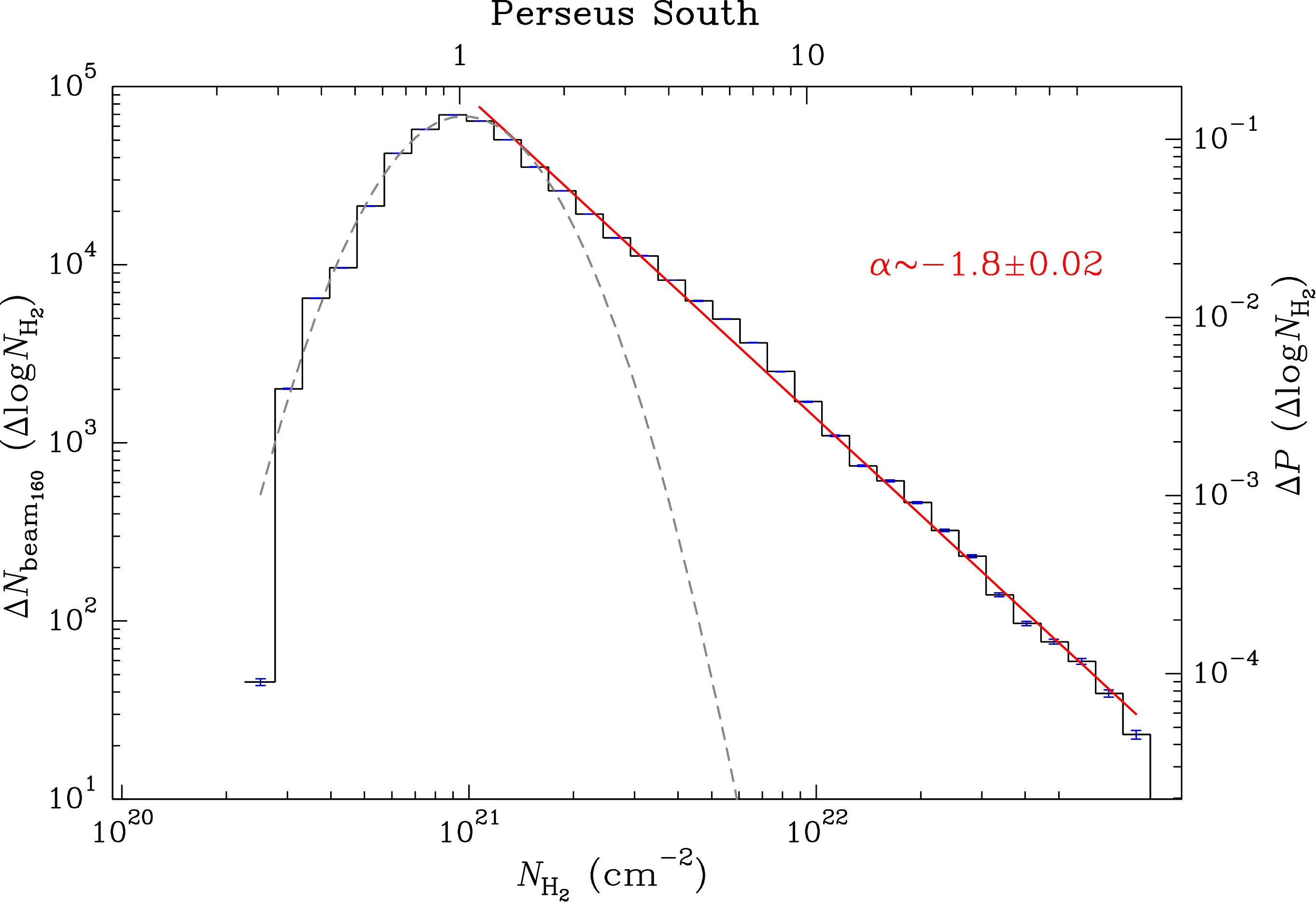}
   \caption{Probability distribution function (PDF) 
   at ${13.5}''$resolution of the column density map in Perseus north and south region.
   The lower axis of the horizontal axis is the column density ($N_{\rm H_{2}}$), 
   the upper axis of the horizontal axis  is the corresponding extinction 
   $A_{V}$ (mag). }
   \label{PDF}
    \end{figure*}

The PDF can be computed 
as the histograms of the column density and can be used to 
characterize physical properties the structure of molecular clouds. 
The PDF of a variable is a one-point statistics that the relative fraction of 
the mass in a given range \citep{VazquezSemadeni+etal+2001}.
The PDF at ${13.5}''$resolution of the column density 
map in Perseus north and south region is shown in Figure~\ref{PDF}.
The $Herschel$ column density converted to visual extinction
units with assumption of $N_{\rm H_{2}}({\rm cm^{-2}})=0.94\times10^{21} A_{\rm v} (\rm mag)$ \citep{Bohlin+etal+2001}.
The PDF can be well fitted with parabola and the peak value at $\sim $1 mag.
But high density region $>$2 mag, Perseus north can be well fitted with a 
power-law with a index of 2.55, and south region can be well fitted with a index of 1.8.
It seems that the power law in the north is steeper than the power law in the south, 
that means that in the same interval, the south side contains more gas than the north side.

\subsubsection{Core selection}
\begin{figure*}
     \centering
     \includegraphics[width=1.0 \textwidth]{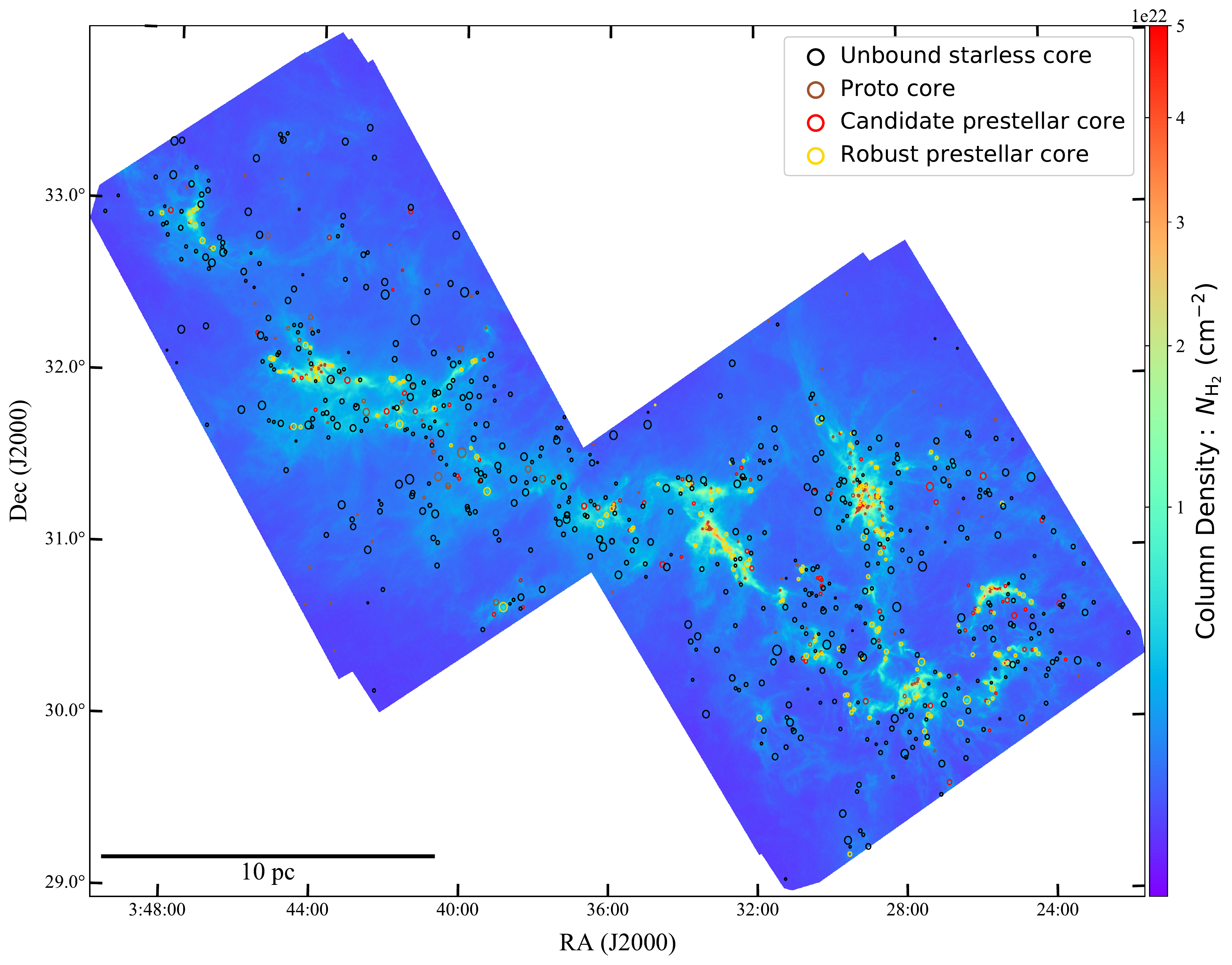}
     \caption{Positions of the 952 dense cores
     identified in the Perseus overlaid on the $Herschel$ high resolution ${13.5}''$
     column density map. Black, brown, red, and yellow circles mark the 536 unbound starless cores,
      the 130 protostellar cores, the 87 candidate prestellar cores, the 199 robust
     prestellar cores, respectively. The size of the circle is the geometric mean of the core's FWHM.}
     \label{coreposition}
        \end{figure*}

\begin{table}
\begin{center}
\caption[]{}\label{Tab1}
 \begin{tabular}{ccccc}
  \hline\noalign{\smallskip}
Linear density & Unbound starless&  Candidate prestellar& Robust prestellar & Protostellar  \\
  \hline\noalign{\smallskip}
$M_{\rm line}$ \textgreater 16           & 57 (10.6\%)  & 26 (29.9\%) & 135 (67.8\%) & 66 (50.8\%)  \\ 
8 \textless $M_{\rm line}$ \textless 16  & 62 (11.6\%)  & 19 (21.9\%) & 36  (18.1\%) & 10 (7.7\%)   \\
$M_{\rm line}$ \textless 8               & 172 (32.1\%) & 28 (32.2\%) & 19  (9.5\%)  & 6  (4.6\%)   \\
Not in filament                               & 245 (45.7\%) & 14 (16.1\%) & 9   (4.5\%)  & 48 (37.0\%)  \\
  \noalign{\smallskip}\hline
\end{tabular}
\end{center}
\end{table}
Sufficiently good cores from multi-wavelength catalogs are selected with criteria below
, which are based on benchmark tests \citep{Men+2021b}. We marked all the cores as circles on the column density map (See Figure~\ref{coreposition}), and four colors of the circles indicate four types of cores. The size of the circles are the geometric mean of core's FWHM.
\begin{enumerate}
\item[--] $\left | \rm GOODM \right |>1$, where $\rm GOODM$ is monochromatic goodness. 
\item[--] $\left | \rm SIGNM \right |>1$, where $\rm SIGNM$ is detection significance from monochromatic single scales. 
\item[--] ${\rm FXP}_{\rm BST}/{\rm FXP}_{\rm ERR}>2$, where ${\rm FXP}_{\rm BST}$ is peak intensity 
and $\rm {FXP}_{ERR}$ is peak intensity error. 
\item [--] $\rm {FXT}_{BST}/{FXT}_{ERR}>2$, where $\rm {FXT}_{BST}$ is total flux 
and $\rm {FXT}_{ERR}$ is total flux error.
\item[--] $\rm AFWHM/BFWHM<2$, where $\rm AFWHM$ is major size at half-maximum and $\rm BFWHM$ is minor size at half-maximum.
\item[--] $\rm FOOA/AFWHM>1.15$, where $\rm FOOA$ is full major axis of an elliptical footprint.  
\end{enumerate}
The cores are classified according to the method described by \citep{Konyves+etal+2015}. 
We briefly outline this method here. 
The integrated flux measured at each wavelength by $getsf$ were used 
to fit a SED with a modified blackbody function, to obtain physical parameters 
such as mass, temperature, and bolometric luminosity of each core.
Prestellar cores are gravitationally bound starless cores most likely 
to form stars \citep{Ward-Thompson+etal+2007, Andre+etal+2014}.
Self-gravitational isothermal equilibrium Bonnor-Ebert (BE) sphere 
is bounded by surrounding gas, similar to the physical state of the prestellar core 
The critical BE mass can be expressed as \citep{Bonnor+1956}
 $M_{\rm BE}^{\rm crit} \approx 2.4\, R_{\rm BE}\, c_{\rm s}^{2}/G$, 
 where $R_{\rm BE}$ is the BE radius, 
 and $G$ is the gravitational constant. Assuming a ambient cloud temperature of 10 K, the 
 isothermal sound speed $c_{\rm s}$ is $\sim 0.2$ km/s.
 We use this model to select prestellar cores.
The core size is defined as the mean deconvolved FWHM diameter at the resolution of ${18.2}''$
of an equivalent elliptical Gaussian source: $R_{\rm dec}=\sqrt{{\rm AFWHM*BFWHM}-{\theta _{{18.2''}}^{2}}}$,
where $\theta _{{18.2}''}$ is the angular resolution at $Herschel$ 250 $\mu $m band.
Assuming an ambient cloud temperature of 10 K, the isothermal sound speed $c_{\rm s}$ is $\sim 0.2 \ \rm km\ s^{-1}$. 
If the ratio $\alpha_{\rm BE}=M_{\rm BE}^{\rm crit}/M_{\rm core}\leq  2$,  
we deem this starless core is self-gravitating and classified as a robust prestellar core. 
\citet{Konyves+etal+2015} propose an empirical size-dependent ratio 
$\alpha_{\rm BE,emp}\leq 5 \times (\sqrt{\rm AFWHM*BFWHM}/\theta _{18.2''})^{0.4}$
is also considered to select candidate prestellar cores. And cores with at least 
one protostar in the half-power column density profile are considered 
to be protostellar cores.
\subsubsection{Statistics of core physical parameters}\label{3.1.3}

\begin{figure*}
  \centering
  \begin{minipage}[t]{1.0\linewidth}
  \centering
     \includegraphics[width=0.49\textwidth]{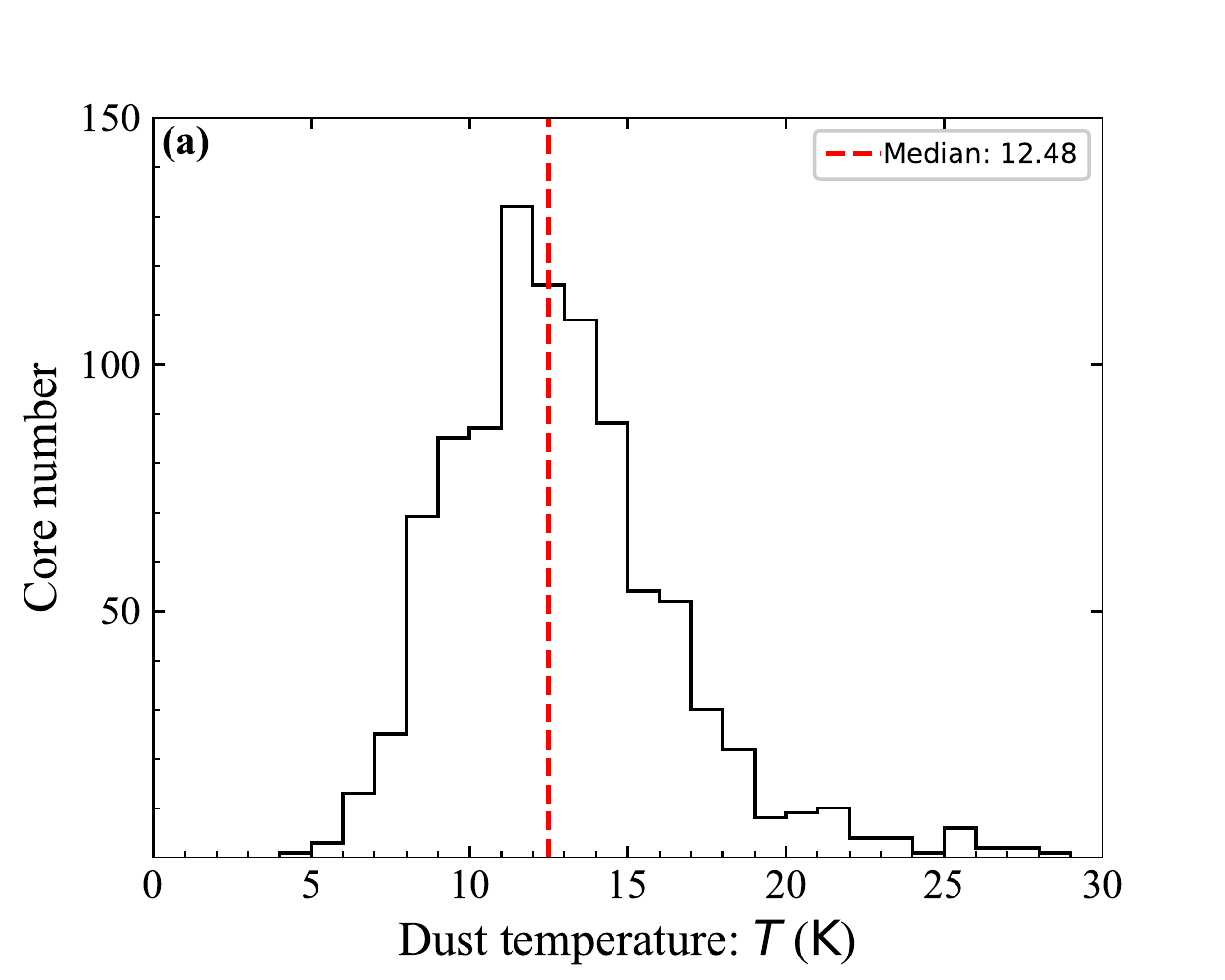}
     \includegraphics[width=0.49\textwidth]{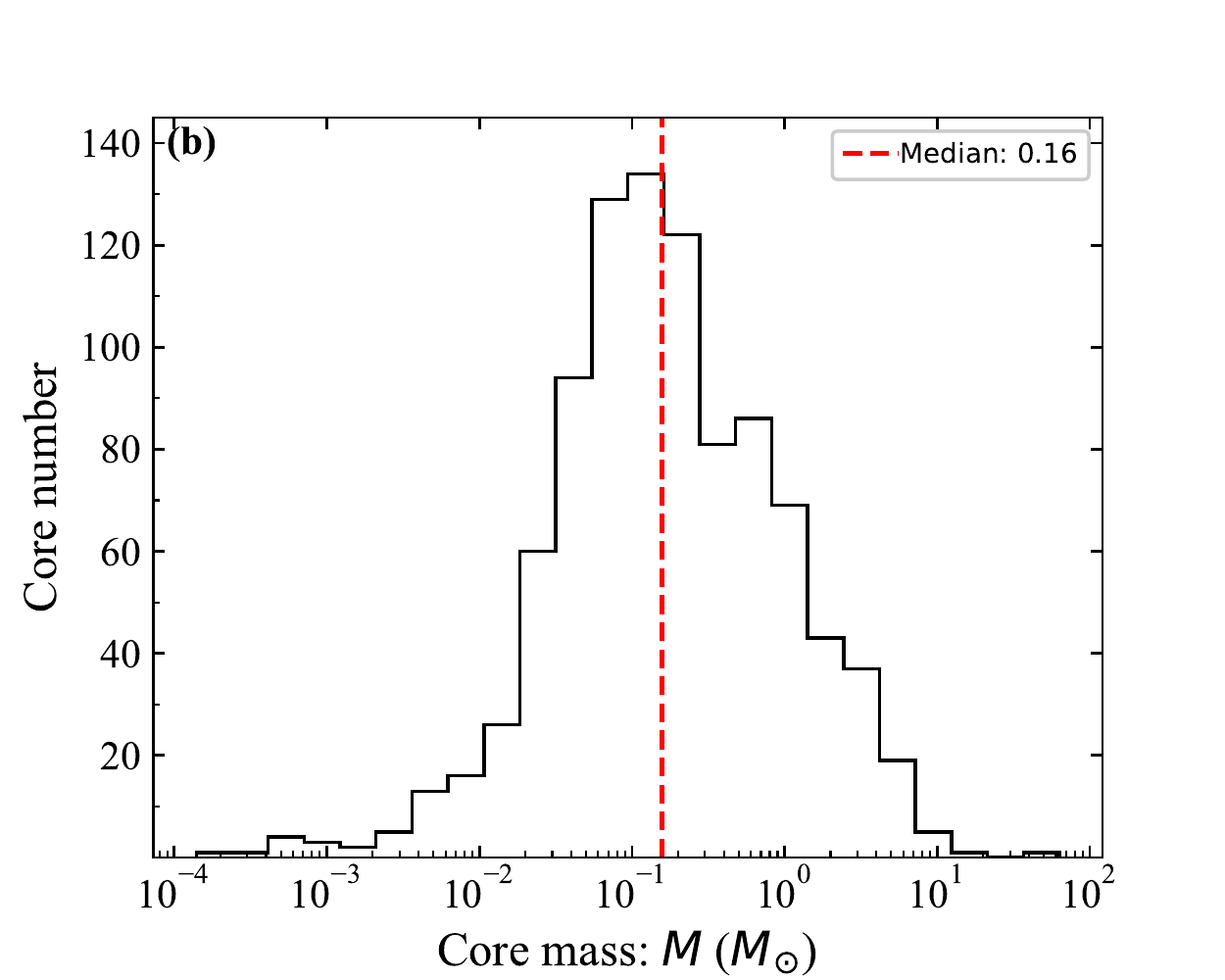}
  \end{minipage}  
  \begin{minipage}[t]{1.0\linewidth}
  \centering
     \includegraphics[width=0.49\textwidth]{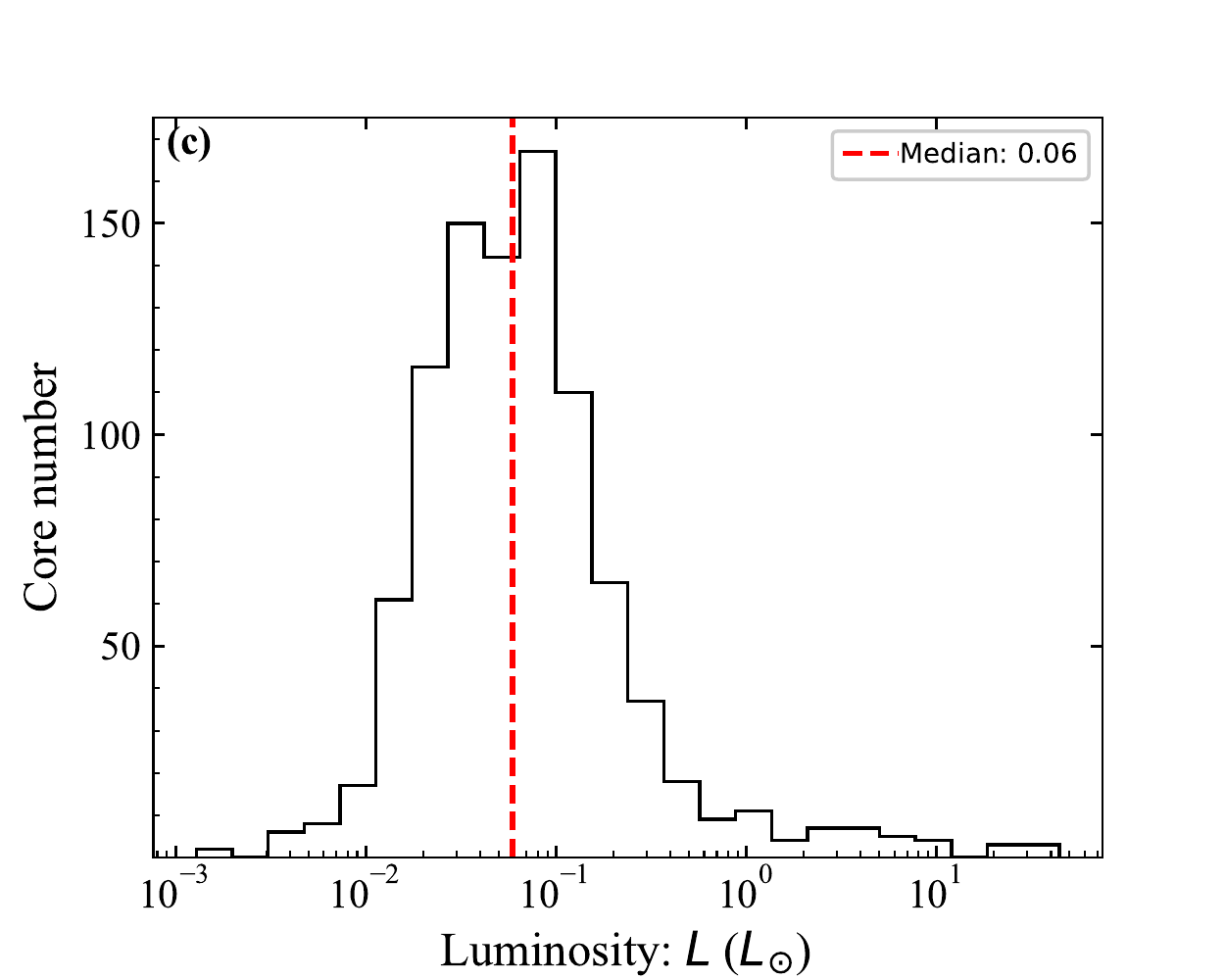}
     \includegraphics[width=0.49\textwidth]{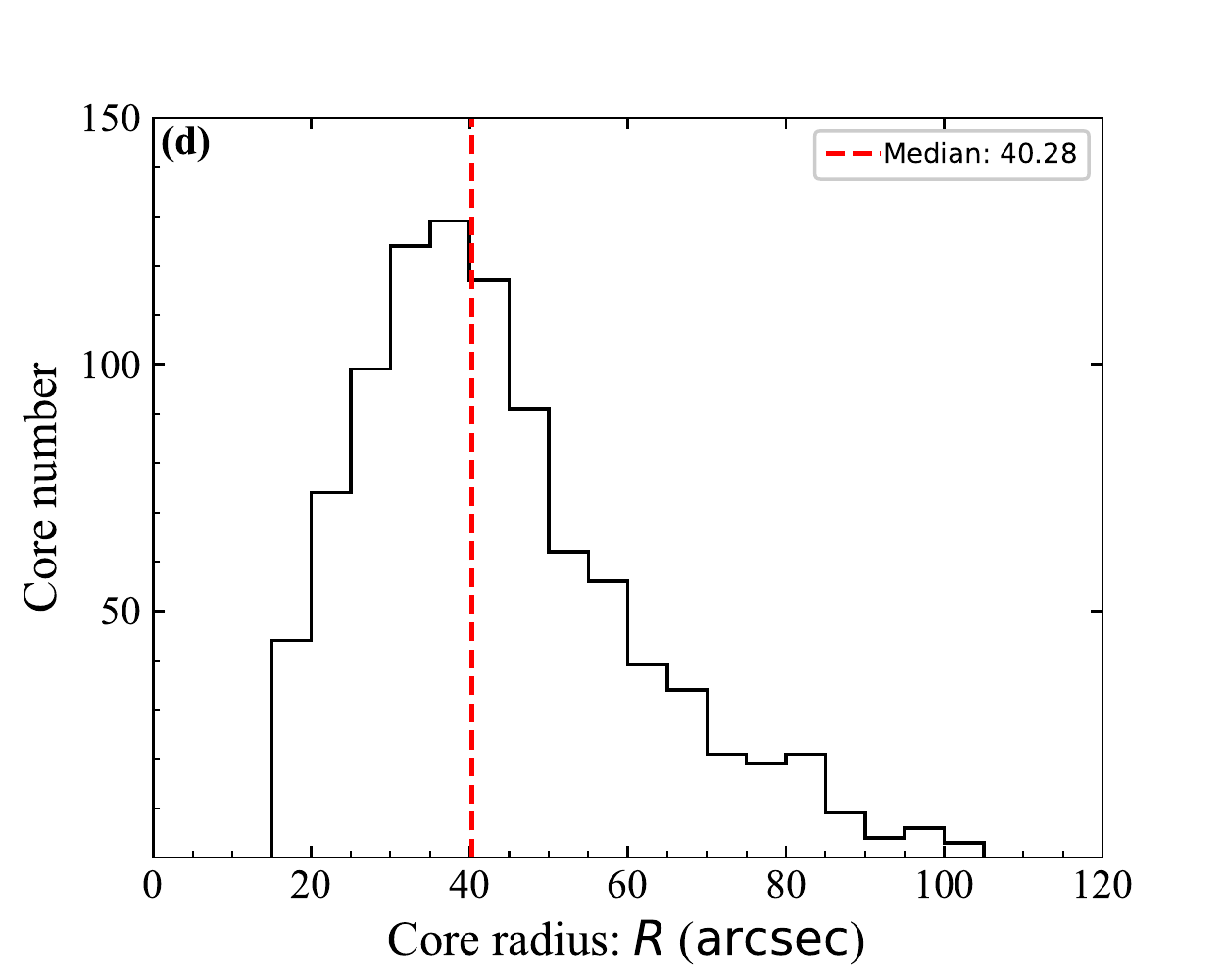}
  \end{minipage} 
     \caption{Histogram of physical parameters of the cores.
     Panel (a), (b), (c): 
     show core dust temperature, mass and bolometric luminosity 
     obtained from SED fitting. Panel (d) is the geometric mean 
     of the core FWHMs. The red dashed lines mark the median value.}
     \label{fig5}
\end{figure*}

\begin{figure*}
  \begin{minipage}[t]{1.0\linewidth}
  \centering
     \includegraphics[width=0.49\textwidth]{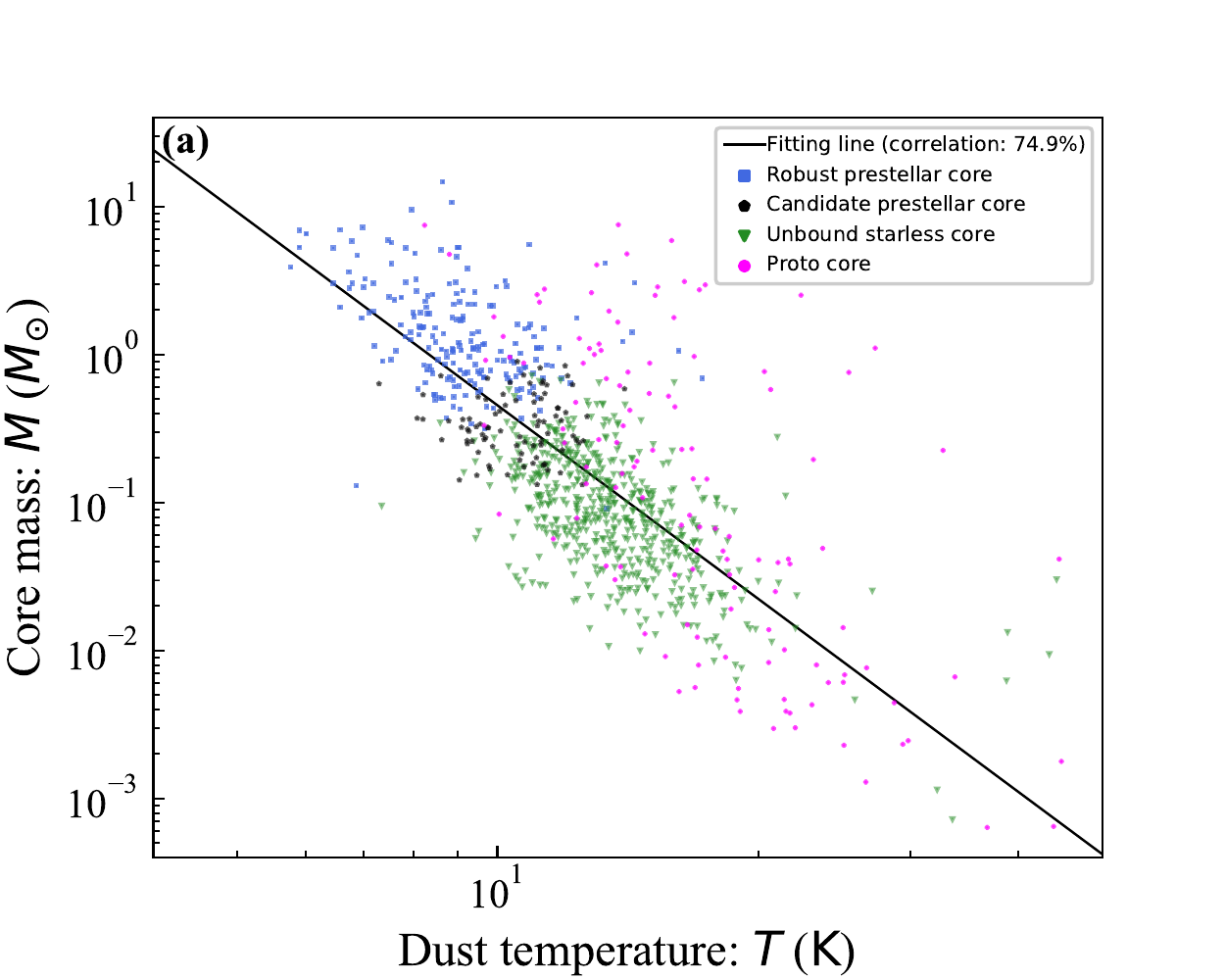}
     \includegraphics[width=0.49\textwidth]{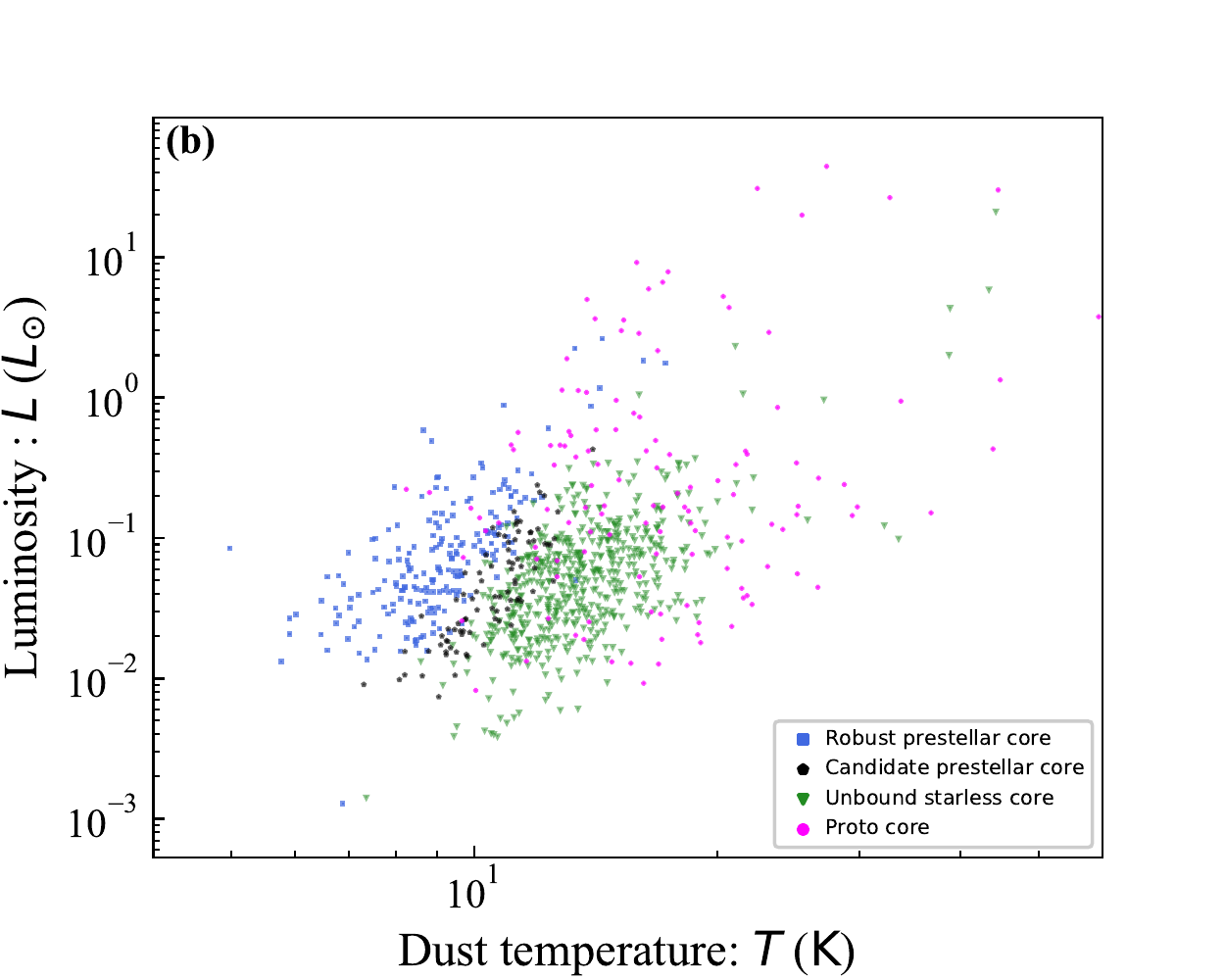}
       \end{minipage} 
  \begin{minipage}[t]{1.0\linewidth}
  \centering
     \includegraphics[width=0.49\textwidth]{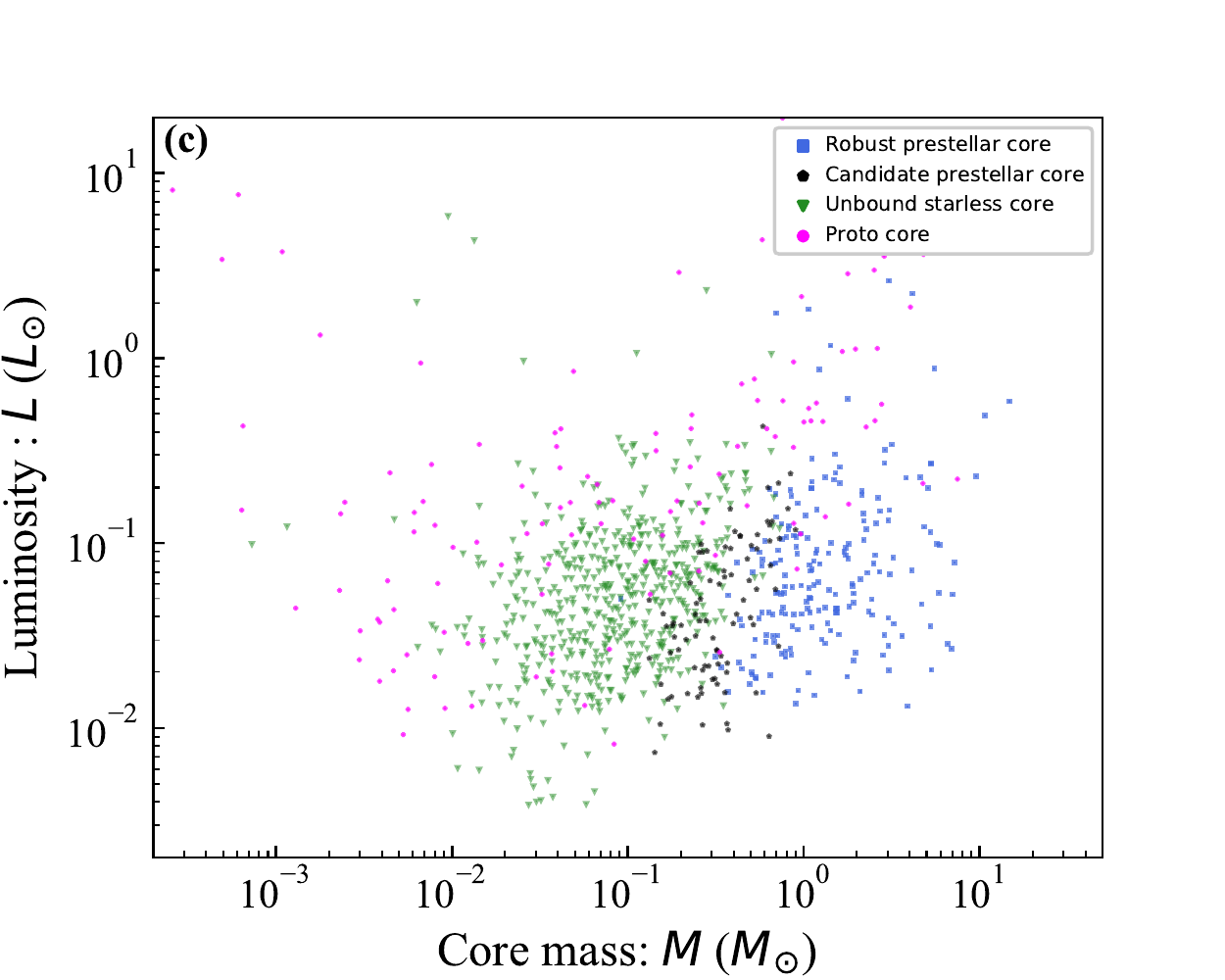}
     \includegraphics[width=0.49\textwidth]{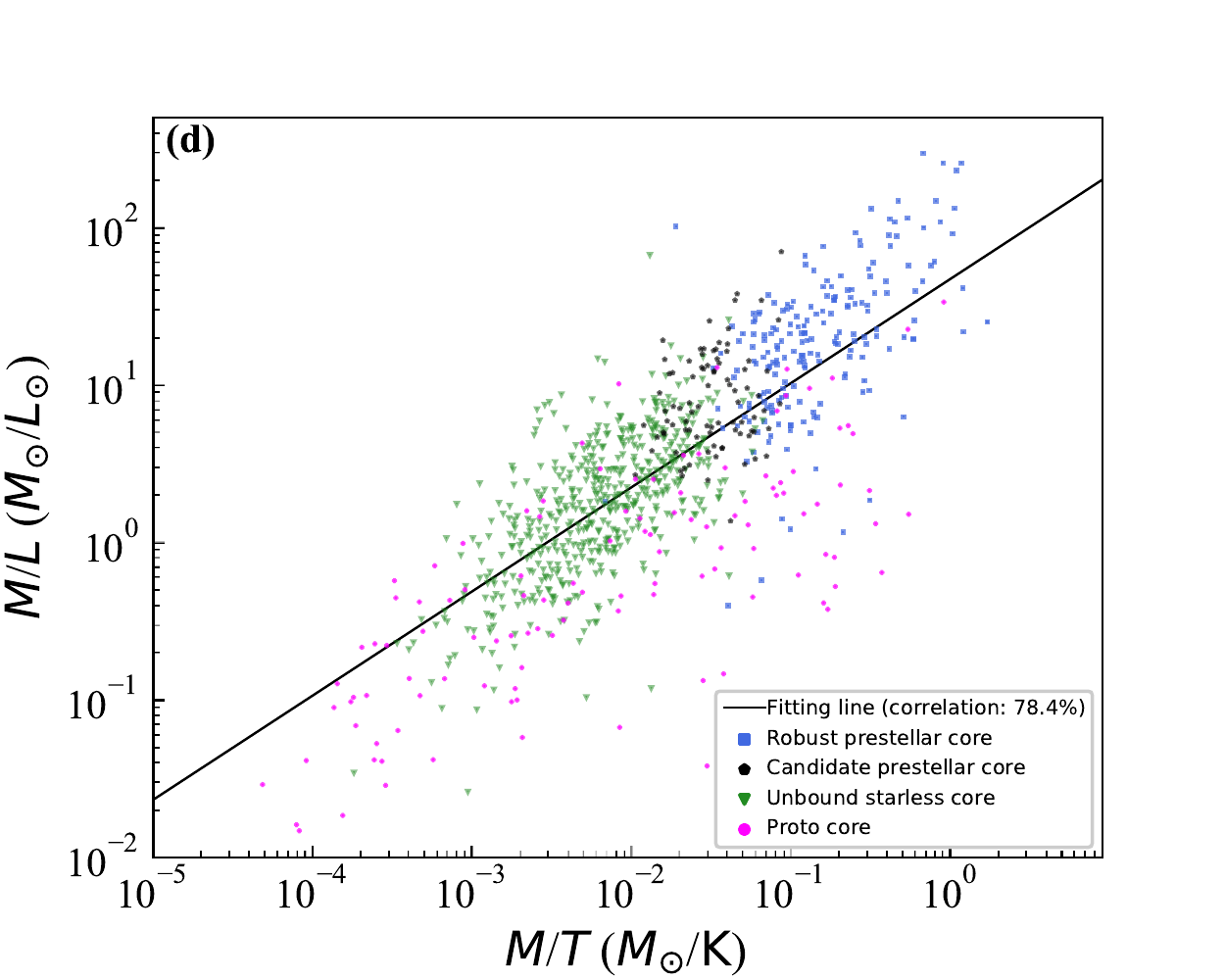}
      \end{minipage}  
     \caption{Correlations among core physical parameters derived by SED fitting.
     Panel (a):  A clear correlation between core dust temperature and mass, 
     ${\rm log}_{10}(M/M_{\odot })=-4.34{\rm log}_{10}(T/{\rm K})+4$
     Panel (b) and (c): Core luminosity has no significant correlation with dust temperature and  mass.
     Panel (d): A clear correlation between $M/L$ and $M/T$, ${\rm log}_{10}(M/T)=0.66{\rm log}_{10}(M/T)+1.67$.}
     \label{fig6}
\end{figure*}

We obtained physical parameters, which include temperature, bolometric luminosity and mass by using SED fitting for each core. Statistical and fitting result of all the cores show in Figure~\ref{fig5} and Figure~\ref{fig6}. Figure~\ref{fig5} shows histogram of core temperature, mass, bolometric luminosity and radius. We obtained 12.48 ${\rm K}$ for median value of temperature, 0.16 ${M_\odot}$ for mass, 0.06 ${L_\odot}$ for bolometric luminosity and 40.28 arcsec for radius.

Figure~\ref{fig6} shows the correlation of each parameter, and blue, black, green and purple markers represent for robust prestellar cores, candidate prestellar cores, unbound starless cores and proto cores, respectively. Figure~\ref{fig6} (a) and (d), show the definitely linear correlation ($R^2_a=74.9\%, R^2_b=79.4\%$) of temperature and mass or ratio of mass to temperature and ratio of mass to luminosity. However, Figure~\ref{fig6} (b) and (c) show both temperature and mass are independent of bolometric luminosity.

\subsection{Filamentary Structure Sample}
\subsubsection{Filamentary Structure Selection}
We use the built-in script $getsf$: $fmeasure$ to measure the filamentary 
structure along its skeleton on the background-subtracted ${13.5}''$ resolution 
column density image of the filament components.
Filamentary structures are three-dimensional structures in space, 
but what $Herschel$ observed are their two-dimensional projections.
Filamentary structures are twist in shape. They blend with themselves 
and with the surrounding structures. 
The column density contrast (C) of the filamentary structure is defined as
$C=N^{\rm crest}_{\rm H_{2}}/N^{\rm bg}_{\rm H_{2}}$, 
where $N^{\rm crest}_{\rm H_{2}}$ is the filament crest column density 
and $N^{\rm bg}_{\rm H_{2}}$ is the filament crest background. 
To choose a clear structure we first select structures with $C>0.5$ from the skeleton network.
This resulted in a sample with 500 segments in the Perseus north region and 
596 segments in the Perseus south region.

The beginning and end of filamentary structures are not easily objectively determined. 
And they have substructures, it is not easy to determine which are substructures 
and which are the main structures.  So as to simplify this complex problem, the 
strategy adopted in $getsf$ by splitting the skeleton network into single segments.
Although, to some extent, the length of the filamentary structures is not objective, 
However, the longer the segment is, the more likely it is to be a filamentary structure, 
which is beyond doubt. In order to further improve the reliability of the sample, 
we select structures with segment length: $L>0.2$ pc from the above sample of $C>0.5$.
This resulted in a sample with 162 segments in the Perseus north region 
and 229 segments in the Perseus south region.

\subsubsection{Statistics of physical parameters of selected filamentary structures}
$fmeasure$ uses two methods to obtain the linear density ($M_{\rm line}$) of 
the filamentary structures. One of the method is that
$fmeasure$ derives the mass ($M_{\rm fil}$) of a filamentary structure by directly 
integrating its footprint and then get $M_{\rm line}$ by $M_{\rm line}=M_{\rm fil}/L_{\rm fil}$, 
where $L_{\rm fil}$ is filament length and filament footprint is defined as the area 
between the skeleton and the largest extent on each side.
Another way is that density integration can be performed at any sampling point along the crest, 
and this integral is $M_{\rm line}$ of this sampling point. We can use $M_{\rm line}$ median value 
of all sampling points as $M_{\rm line}$ of this filamentary structure. 

\begin{figure*}
     \centering
     \includegraphics[width=0.9\textwidth]{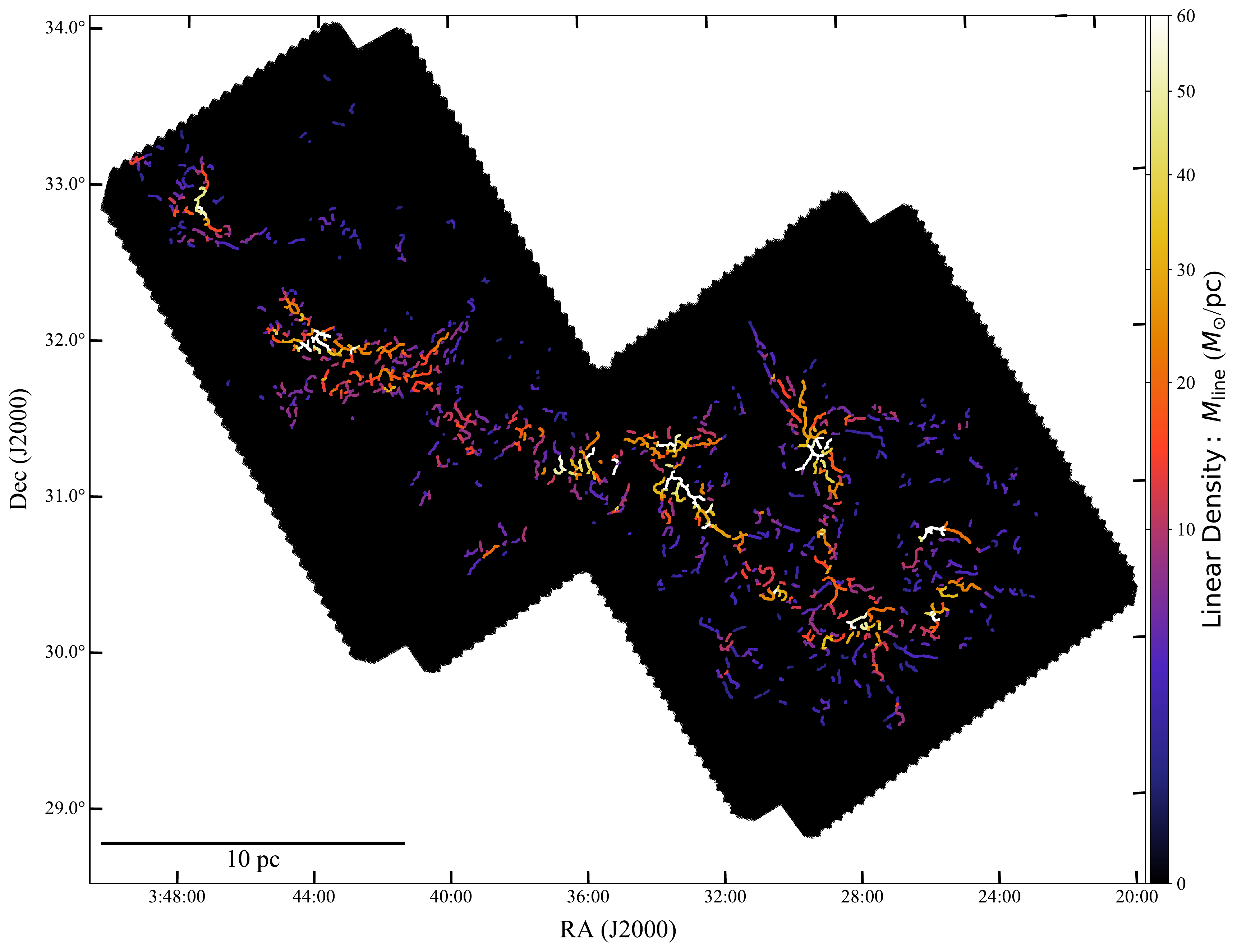}
     \caption{
     Perseus filamentary network structures selected with contrast $C>0.5$. 
     The linear density ($M_{\rm line}$) of the filamentary structure is only taken on the 
     narrow side to avoid contamination as much as possible.}
     \label{fig7}
        \end{figure*}

\begin{figure*}[t]
  \begin{minipage}[t]{1.0\linewidth}
  \centering
     \includegraphics[width=0.49\textwidth]{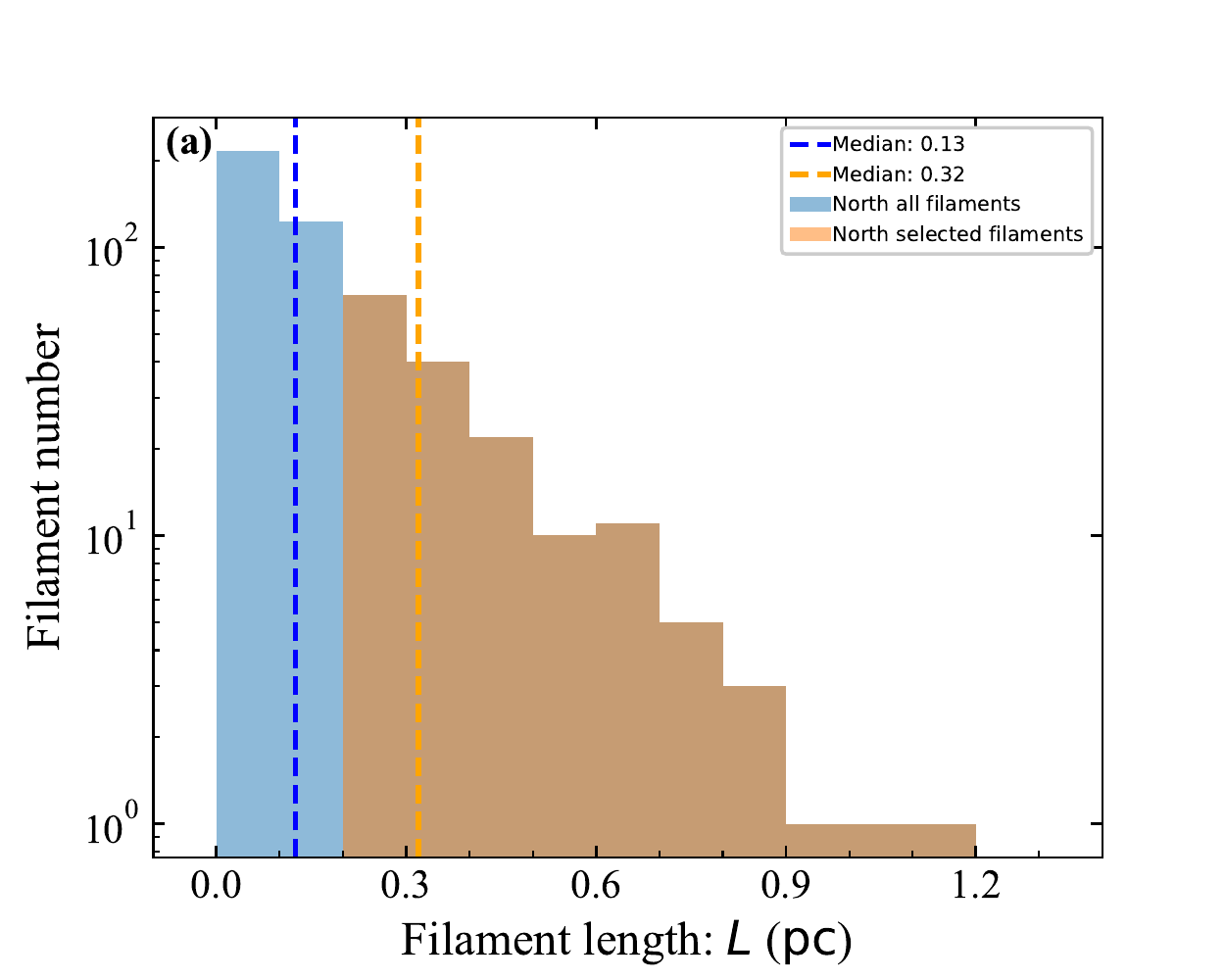}
     \includegraphics[width=0.49\textwidth]{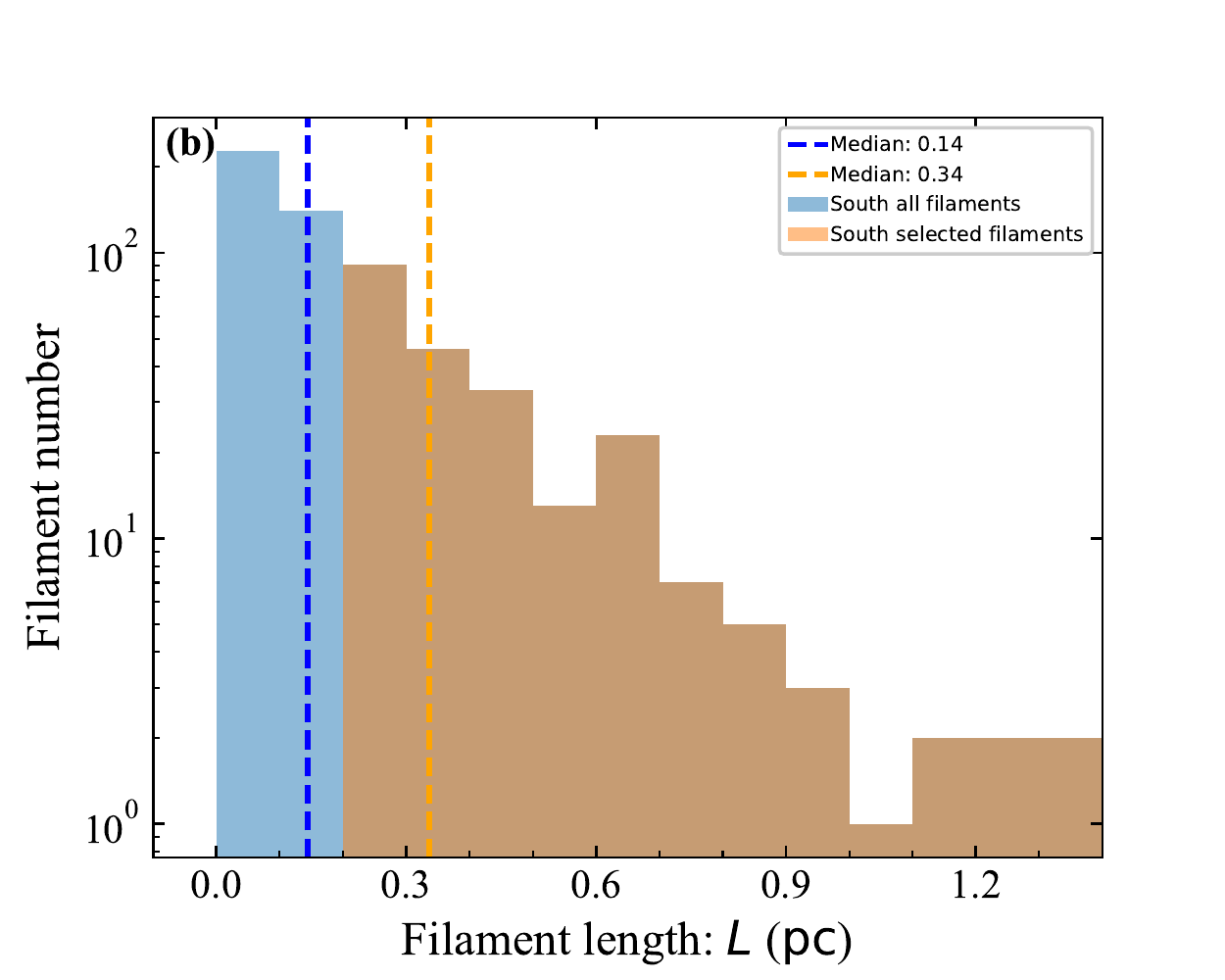}
   \end{minipage}
  \begin{minipage}[t]{1.0\linewidth}
  \centering
     \includegraphics[width=0.49\textwidth]{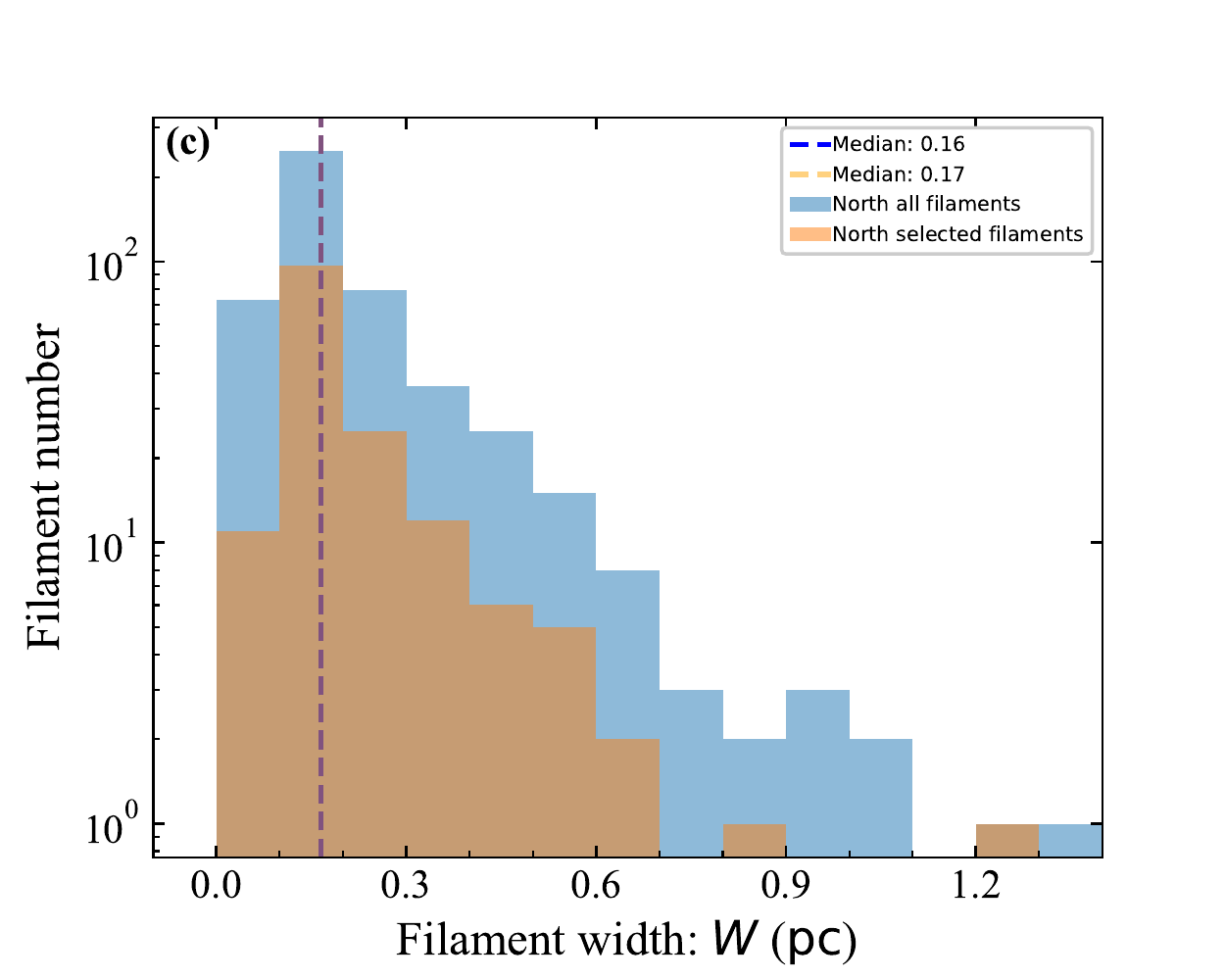}
     \includegraphics[width=0.49\textwidth]{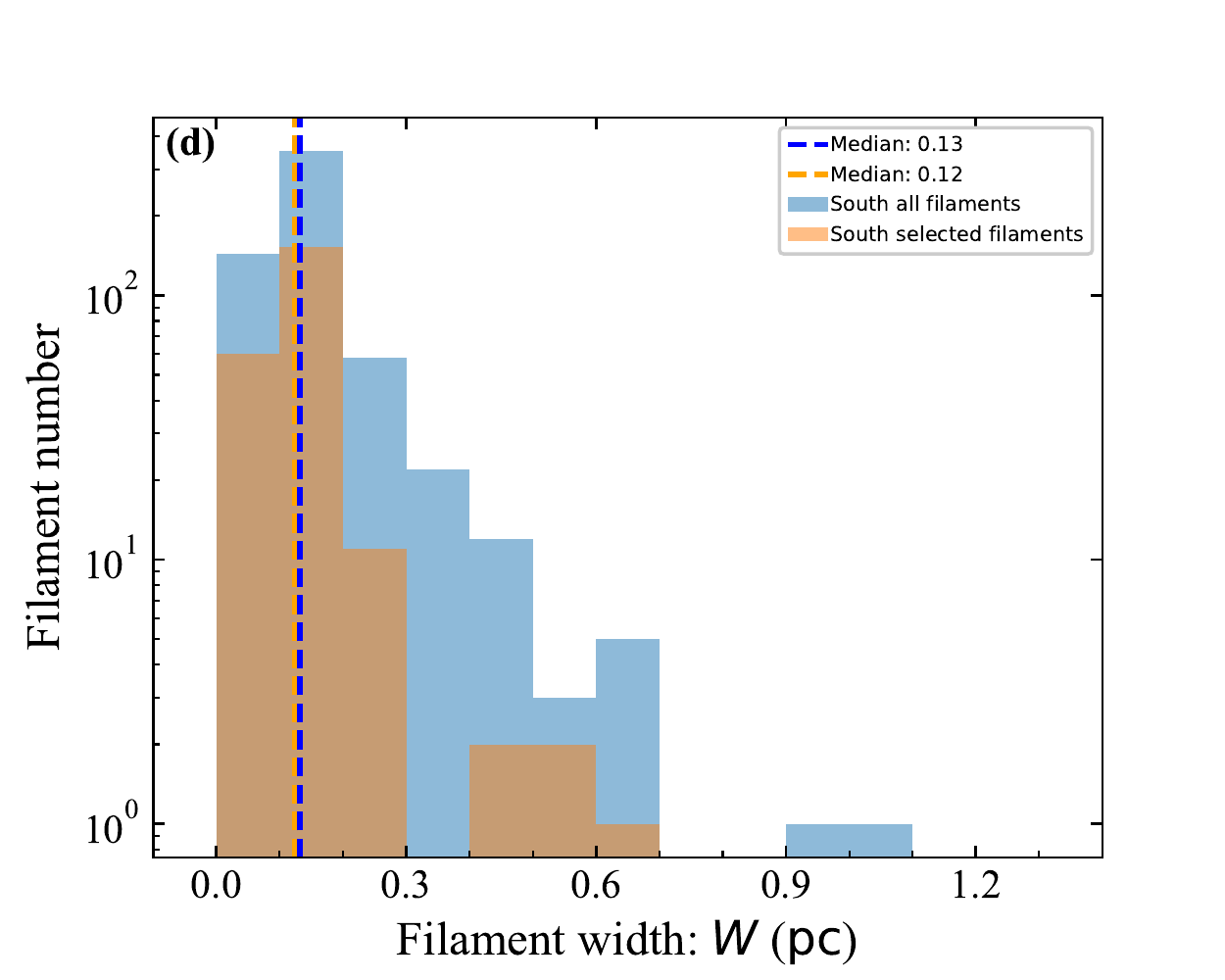}
  \end{minipage}
  \begin{minipage}[t]{1.0\linewidth}
  \centering
     \includegraphics[width=0.49\textwidth]{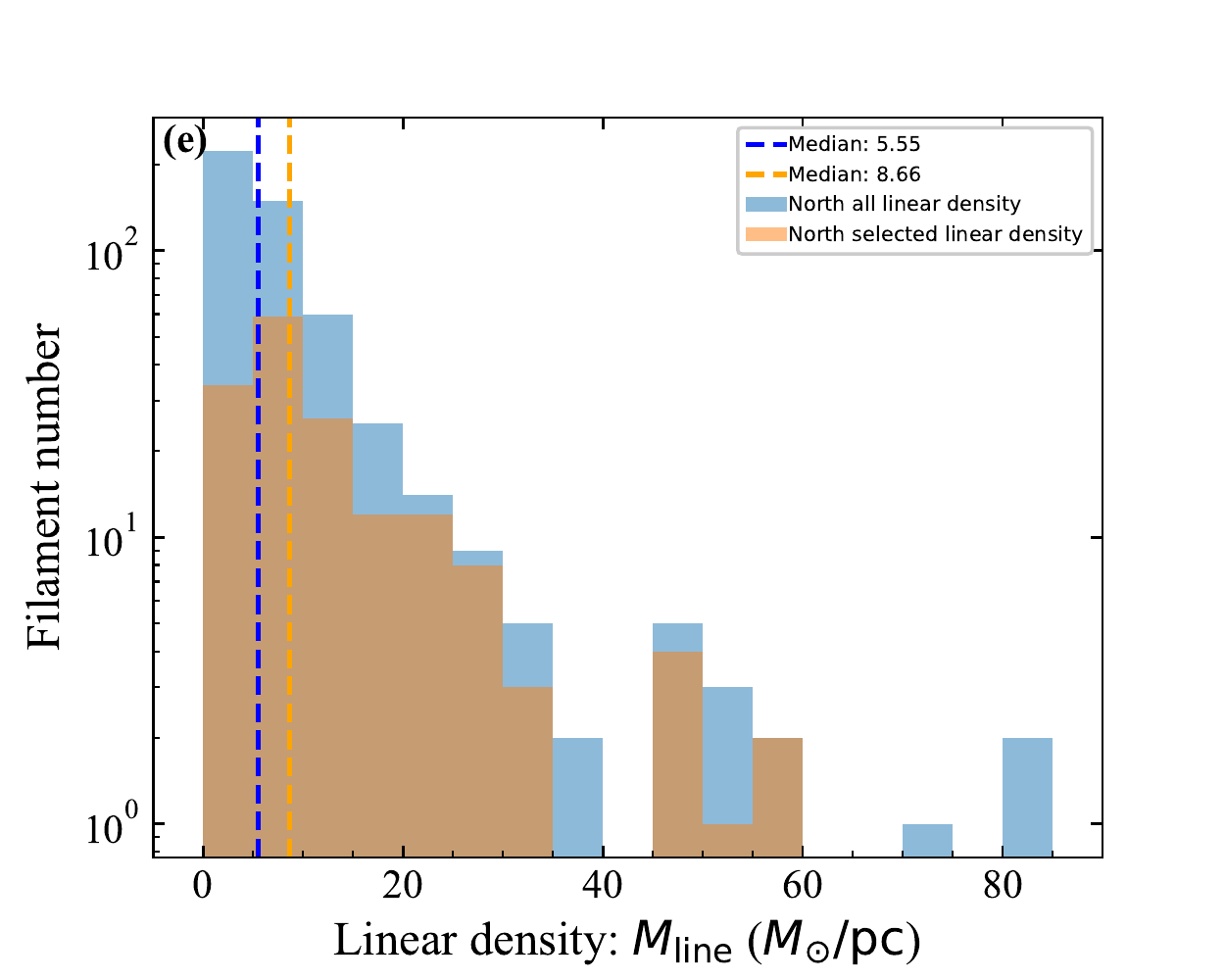}
     \includegraphics[width=0.49\textwidth]{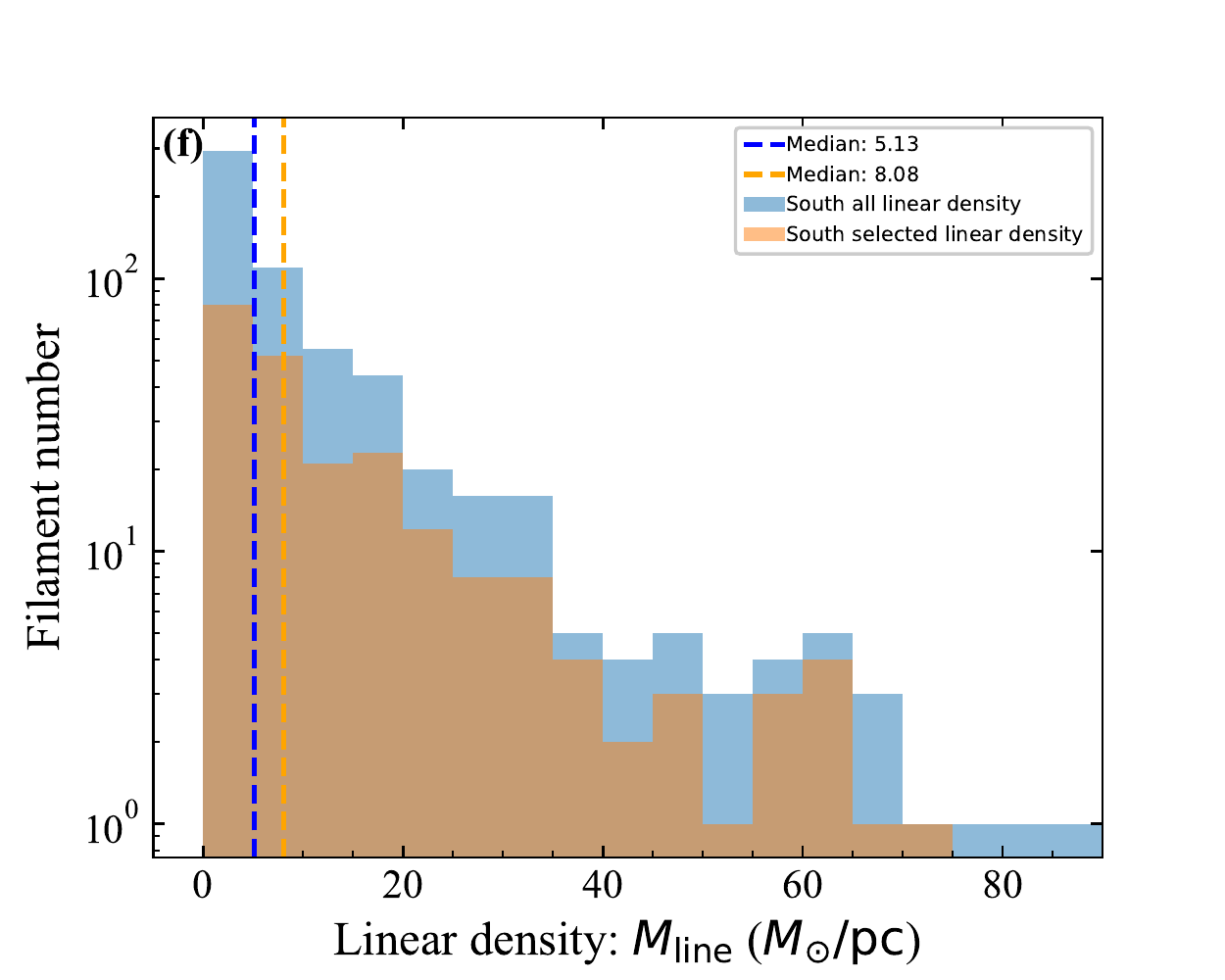}
   \end{minipage}
     \caption{Histogram of physical parameters of the filamentary structures.
     In this set of histograms, 
     the sample "all" are the filamentary structures with $C>0.5$;
     the sample "selected" are the filamentary structures with $C>0.5$ and $L>0.2\rm\ pc$.
     And we conduct separate statistics for filamentary 
     structure in the north and south regions.}
     \label{fig8}
\end{figure*}

\begin{figure*}[t]
  \begin{minipage}[t]{1.00\linewidth}
  \centering
     \includegraphics[width=0.49\textwidth]{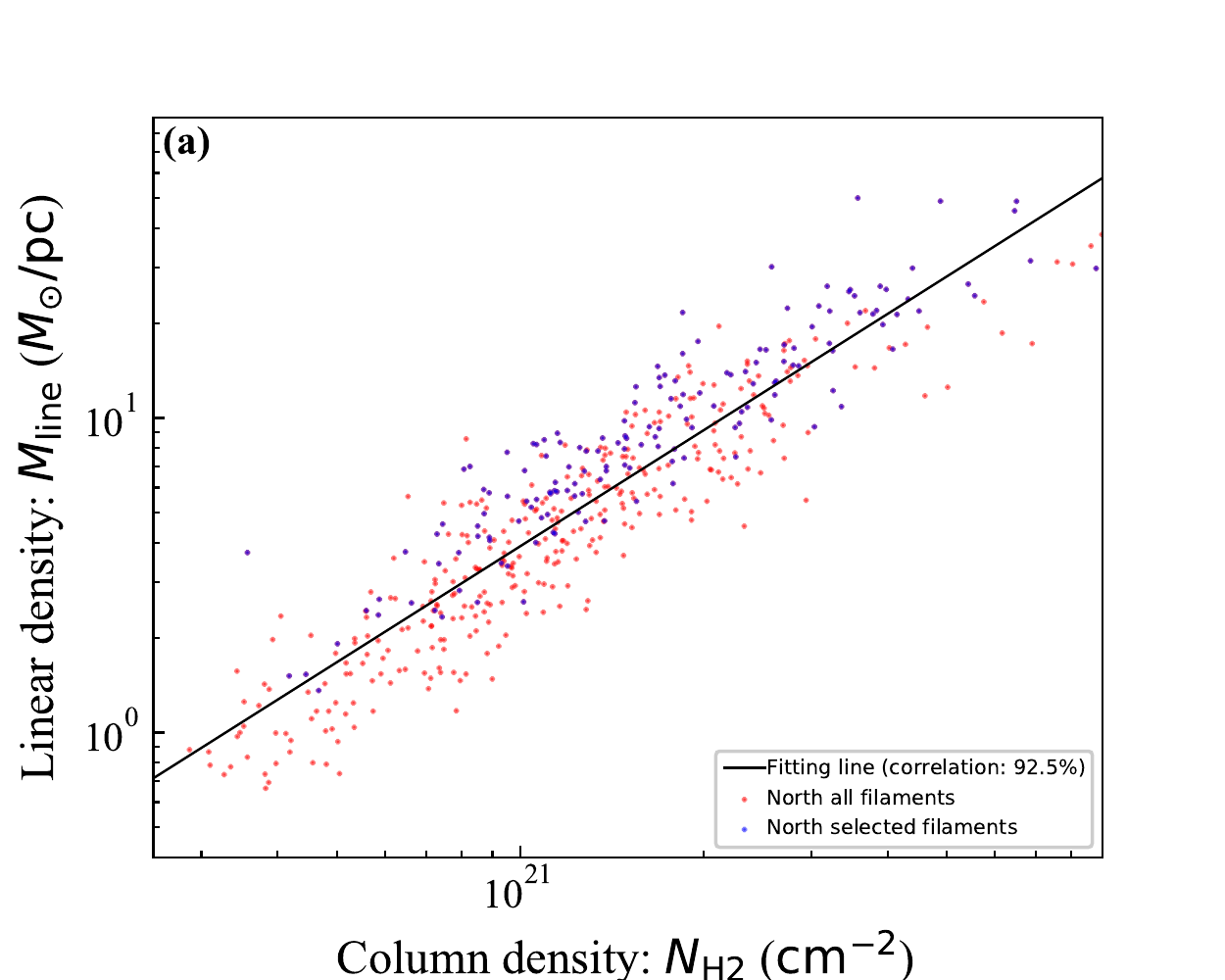}
     \includegraphics[width=0.49\textwidth]{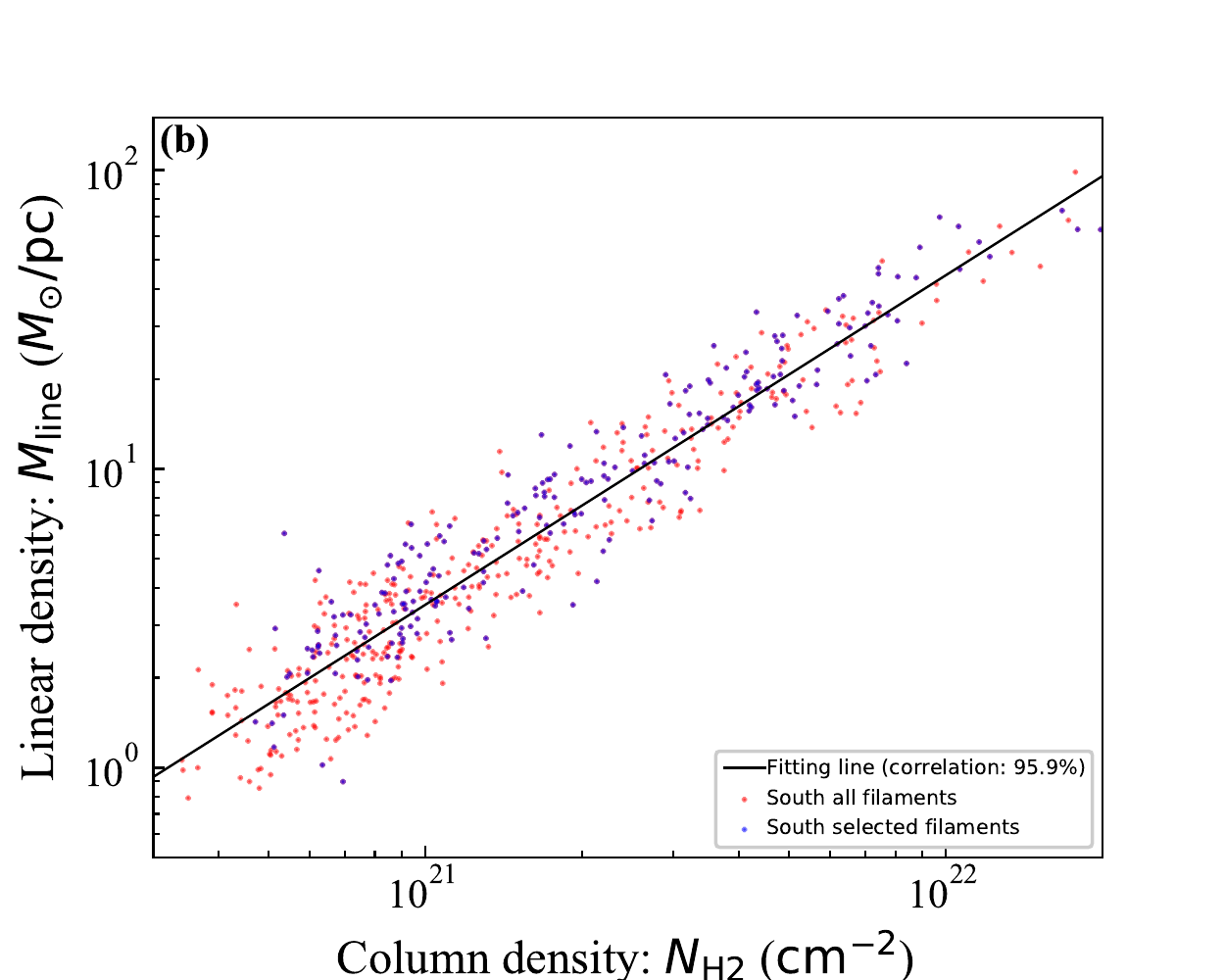}
  \end{minipage}  
  \begin{minipage}[t]{1.00\linewidth}
  \centering
     \includegraphics[width=0.49\textwidth]{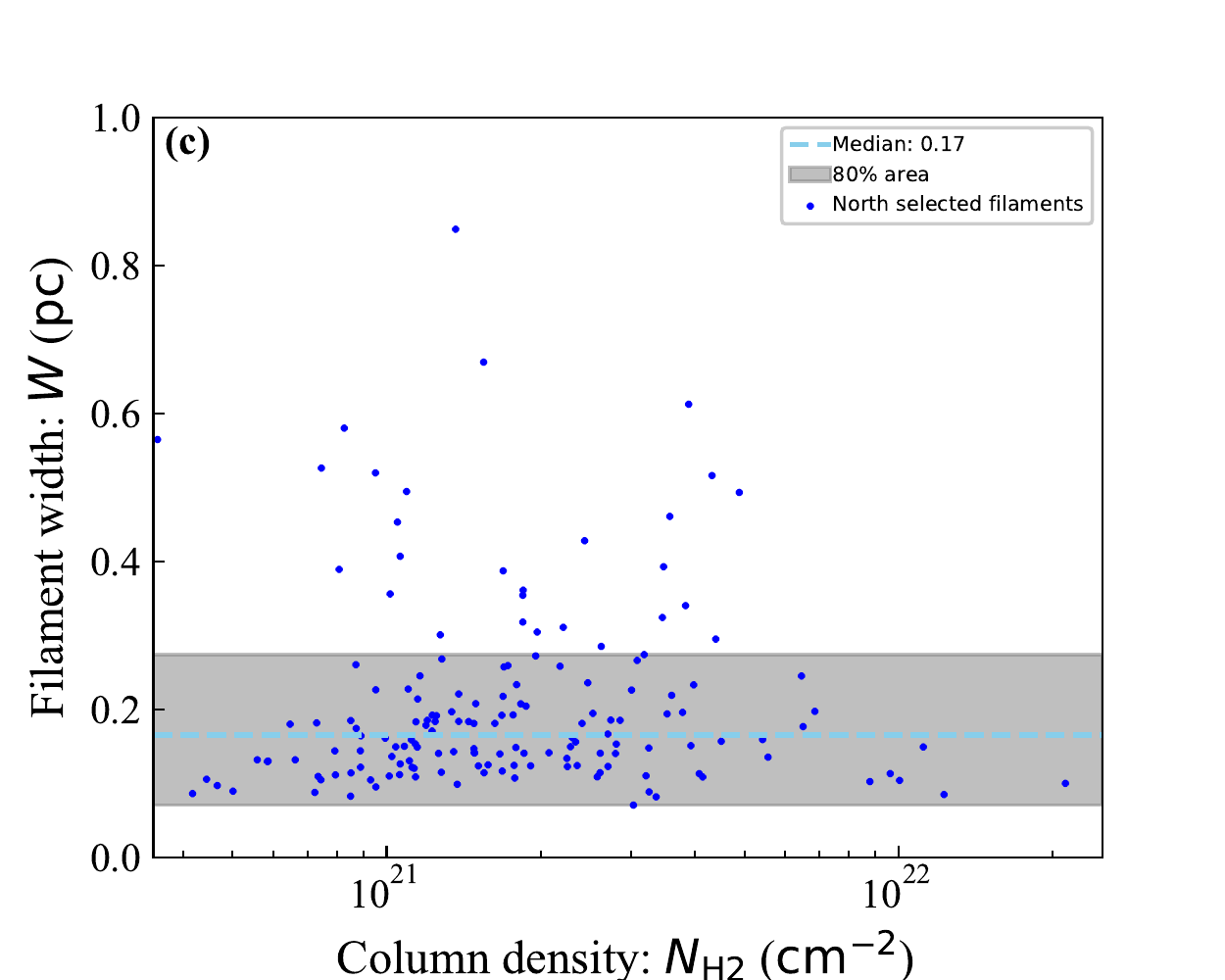}
     \includegraphics[width=0.49\textwidth]{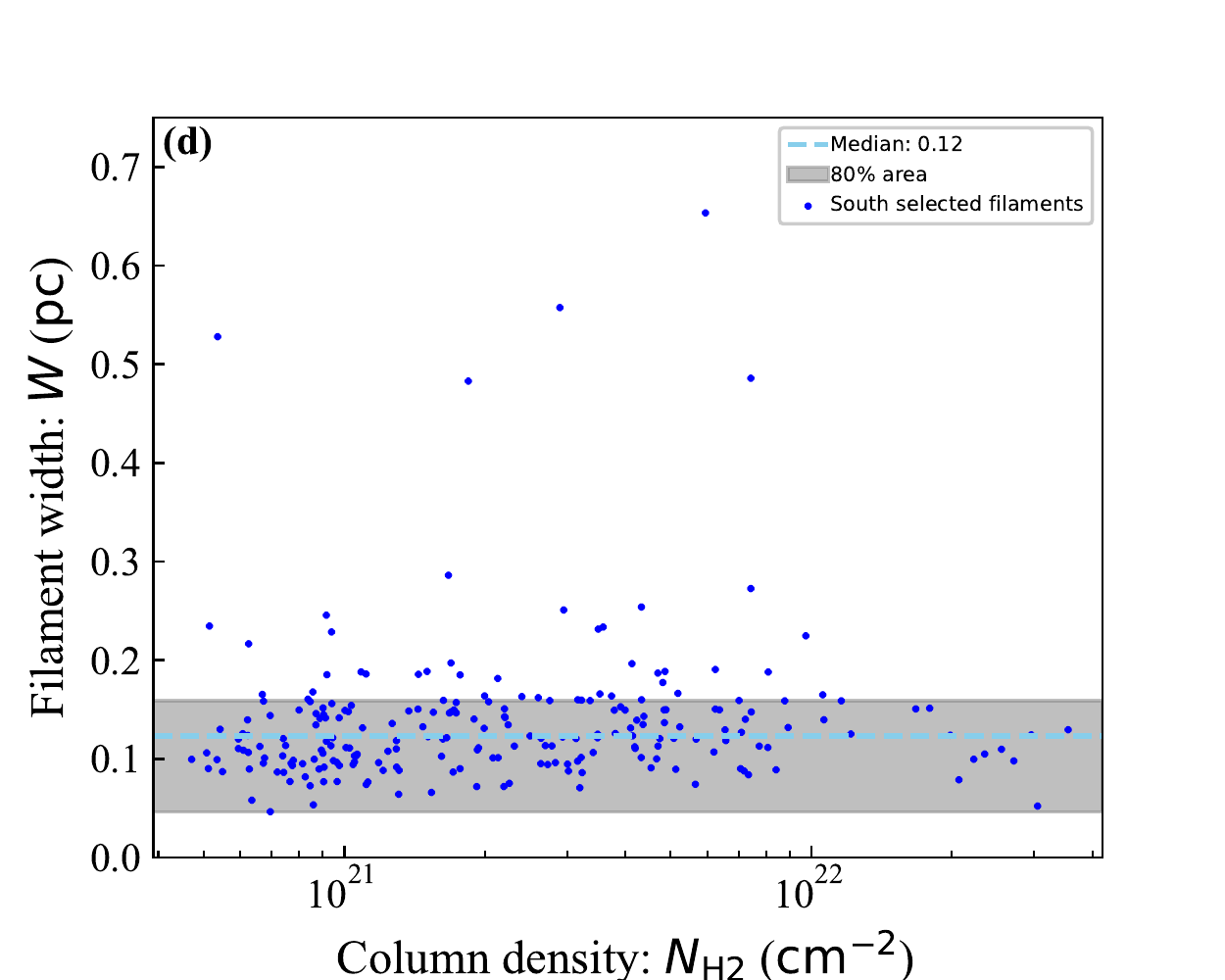}
  \end{minipage}  
     \caption{Crest column density ($N_{\rm H_{2}}$) of filamentary 
     structures as a function of linear density ($M_{\rm line}$) and 
     width ($W$) in the north and south regions of Perseus molecular 
     cloud. $N_{\rm H_{2}}$ and $M_{\rm line}$ are clearly related.
     $N_{\rm H_{2}}$ are uncorrelated with $W$ by three orders of 
     magnitude from $\sim 10^{20}$ to $\sim 10^{22}$ $\rm cm^{-2}$.}
     \label{fig9}
\end{figure*}

We drew  all the filamentary network structure selected with contrast $C~>$ 0.5 in Figure~\ref{fig7}, and the color of the filaments represent the intensity of linear density. The filament of whiter the color means higher linear density. In Figure~\ref{fig8}, we counted the parameters of the filaments in the north and south regions respectively, and made histograms, in which the blue bars represent all the filaments and orange bars represent high reliable filaments. Figure~\ref{fig8} (a) (c) (e) show the length, width and linear density of filaments in north region, then (b) (d) (f) show the parameters in south region. From the median value, the length of filaments in the south region is greater than that of the north region and the width of the filaments contrast between the north and south region is just the opposite. Although the linear density of filaments in the south is slightly lower than that in the north, it is obvious that the supercritical ($M_{\rm line}>16~M_{\odot}/{\rm pc}$) filaments are more distributed in the south region. Figure~\ref{fig9} shows the correlation of the filaments between column and linear density or column density and width. Figure~\ref{fig9} (a) (b) suggest clear correlation (${R_a^2=92.5\%,R_b^2=95.9\%}$) of column density and linear density whether in the South or the north region. However (c) (d) show that column density are uncorrelated with width, and most (80\%) width of filaments are distributed in gray areas. The blue dotted lines represent the median value, with 0.17 pc and 0.12 pc on the north and south region. 

\section{discussion}
\label{sec:discussion}
\subsection{The evolution stages of the north and south are different.}
The thickness of the molecular cloud traced by different probes is different in the direction of sight.
This results in different density distributions detected by different probes of molecular cloud. 
But in general, there is a the rapidly decreasing trend of PDF at the high density range.
This can deduced that the dense structures occupy only a small fraction of the volume of 
the molecular clouds and most of the volume is filled with low density background gas.

For isothermal, supersonic, turbulent gases, shock produces a random density enhancement 
proportional to the mean density. According to the central limit theorem, the PDF is a 
lognormal function \citep{VazquezSemadeni+1994}. PDF should be made only for those 
contours that are closed in the map. In numerical simulation, cloud boundaries are 
generally equivalent to non-closed density contours in numerical boxes. The range of 
these column densities is underestimated, resulting in spurious drops in the PDF.
The PDF of the molecular cloud does not necessarily decrease at low densities, 
and inferring the physical properties of the molecular cloud by fitting the 
low-density region of the PDF may be wrong. Therefore, great care should be 
taken when extrapolating the PDF shape of the cloud in the numerical simulation 
to the real observation data \citep{Alves+etal+2017}. The background component of 
the molecular cloud that exhibits a log-normal function is uncorrelated with 
specific star formation activity, however The PDF dense component exhibits 
a power-law distribution that is closely related to star formation \citep{Kainulainen+etal+2009}.
Numerical simulations predict that turbulence dominates the lognormal distribution, 
while gravity leads to a power-law form \citep{Kritsuk+etal+2011}.
The Perseus south region shows a power law with an exponent of -1.8. 
The Perseus north region has an exponent of -2.55.  
The Perseus south region is flatter than Perseus north region 
and closer to power law exponent of -1.35 for 
stellar initial mass function (IMF) proposed by \citet{Salpeter+1955}. 

The gravitational potential energy is $\Omega =-\alpha \frac{GM^{2}}{R}$, where $\alpha$ is fudge factor of order unity 
that depends on the internal density structure. 
$M$ is the observed structure mass,and $R$ is the effective radius. 
The 3$\sigma$ noise level is $\sim 2\times 10^{21} \ \rm cm^{-2}$, 
which was estimated at regions without sources.
The masses of gas component with $N_{\rm H_{2}}>3\sigma$  
are $\sim 1.8\times 10^{3}\ M_{\odot}$ and $\sim 3.3 \times 10^{3}\ M_{\odot}$
at north and south region respectively, and the effective radii are 2.8 pc 
and 3.2 pc. The $\Omega$ at north region is 
$1.2\times 10^{6}\alpha G \ {M_{\odot }}^{2} {\rm pc}^{-1}$, 
and the $\Omega$ at south region is $3.4 \times 10^{6}\alpha 
G \ {M_{\odot }}^{2} {\rm pc}^{-1}$.
Assuming that south and north region have the same $\alpha$ value, 
the gravitational force of Perseus in south region is 
about three times stronger that in the north. 

Perseus MC is a medium active star forming region.
\citep{Mercimek+etal+2017} identified 222 YSOs in Perseus MC.
Number of YSOs per unit area that is similar to NGC 2264,
but higher than Orion B and lower than Ophiuchus \citep{Pokhrel+etal+2020}.
%There are only 222 identified YSOs \citep{Mercimek+etal+2017}, and 
The number of YSOs (153) in south region is 
significantly more than that in the north region (69). 
Most of the YSOs in the south region are class II/III, and 
the evolution stage is obviously later than that in the north region. 

Because the gas and dust will block the photon radiation, 
the denser region of gas and dust in the molecular cloud will show lower temperature. 
We checked the temperature map officially released by Planck and found that 
there is a clear temperature difference between the north and south regions.
The dust temperature of large-scale background gas of the north 
 region is at 18 K. Dense gas components ($N_{\rm H_{2}}>3\sigma$) are closely 
 related to star formation. IC 348 Reflection Nebula reaches 20-24 K, 
 which covers 70\% of the area of dense gas component in the north region.
 %but dust temperature of B2 and B1-E is about 17 K. 
 The dust temperature of large-scale background gas of the south 
 region is at 17 K. Dust temperature of dense gas component is 
 about 14-15 K in the south region. Dust temperature as an indicator 
 to characterize the difference between the two regions, which 
 have a temperature difference of about 5 K in two regions for 
 these dense gas components.
To conclude, differences in PDF power law exponent of the PDF, 
gravity, number of YSOs, and dust temperature
indicate that the evolution stages of the  north and south regions of Perseus MC are different. 
\subsection{Core evolution}
Core dust temperatures strongly correlated with masses 
and $M/L$ strongly correlated with $M/T$. 
Our results are similar to those of previous studies, 
Such as \citet{Marsh+etal+2014} and \citet{Marsh2016} show that central temperature  
linearly negatively correlated with masses of starless 
and prestellar cores in Taurus, and confirmed a more 
stronger correlation with peak column density.
An intuitive explanation is that as the strength of the shielding core from 
interstellar radiation field increases, the temperature will decrease and the 
mass will increase, in theory, which may be explained by blackbody radiation.  Assuming dust radiation is optically thin at a certain frequency
the measured fluxes can be written as:
\begin{equation}
f_{\nu} = B_{\nu}(T) \,\kappa_{0}\left(\nu/\nu_{0}\right)^{\,\beta} \eta M D^{-2}.
\end{equation}
For a core at a distance $D$ and measured flux $f_{\nu }$ at a certain frequency $\nu $, 
the luminosity at this $\nu $ is 
\begin{equation}
L_{\nu }=4\pi D^{2} f_{\nu }
\end{equation}
Bolometric Luminosity $L_{\rm bol}$ is the luminosity of a core measured 
over all frequency, which is derived by: 

\begin{equation}
L_{\rm bol} =4\pi D^{2} \int f_{\nu }d\nu
\end{equation}
Then 
\begin{equation}
M/L_{\rm bol}\sim \frac{M}{D^2*\int f_{\nu }d\nu}
\end{equation}
And 
\begin{equation}
\int f_{\nu }d\nu \sim {\nu}_{\rm peak}*{\rm SED}_{\rm peak}
\end{equation}
Then according to Wien's law: ${\nu}_{\rm peak}\sim T$,
we can deduce that $M/L_{\rm bol}\sim M/T$

\subsection{Physical properties of ubiquitous filamentary structures}

$Herschel$ observations show that filamentary structure are indeed 
ubiquitous in the molecular cloud \citep[e.g.][]{Hill+etal+2011,Men+etal+2010} 
In the nearby clouds ($<500 $\ pc), filaments profiles measured on $Herschel$ column 
density map in the radial direction, show a typical inner width $\sim $ 0.1 pc and 
no wider than $\sim $0.2\ pc \citep{Arzoumanian+etal+2011,Arzoumanian+etal+2019}.
The origin of the typical inner width of filamentary structures remains a controversial topic.
There are currently three explanations for typical inner width.
The gravitational and thermal pressure equilibrium of the isothermal 
gas results in this typical inner width of 0.1 pc with a weak dependence on 
column density. Typical inner width is just a result of the mechanical equilibrium in 
thermodynamics in radial direction \citep{FischeraandMartin+2012}.
An alternative explanation is that the filaments originate from plane-intersecting 
shock waves due to supersonic interstellar turbulence, and that the filament width 
corresponds to the (magneto-)sonic scale \citep{PudritzandKevlahan+2013}.
Finally, another possible explanation is that the typical inner width of 
the filament may be set by a dissipation mechanism Magnetohydrodynamic (MHD) waves 
induced by ion-neutral friction \citep{Hennebelle+2013}.
The width measured with selected filamentary structure with $L>0.2$ pc and $C>0.5$, 
has higher confidence than the entire filamentary network.
This sample includes 162 segments in the Perseus north region and 229 
segments in the south region. 
The width median value in north region is 0.17 pc and 0.12 pc in south region.  
Large-scale diffuse gas is more abundant in the north than in the south. 
The blending in the north is more severe, that the measured width is 
wider than the south. The measurement of width is consistent with 
the typical inner width of the filamentary structure is 0.1 pc 
measured by \citep{Arzoumanian+etal+2011,Arzoumanian+etal+2019}.

Supercritical filamentary structures play an important role in star formation.
Stars are formed in molecular filaments with linear masses equal to or greater 
than the critical linear mass \citep[e.g.][]{Andre+etal+2014}.
The critical linear mass of an isothermal cylindrical filamentary 
structure depends on temperature and mean molecular weight. Mean molecular weight 
is in turn depends on the metallicity, 
which in turn can have a dependence on the location in the Galaxy.
If we adopt 10 K of ambient cloud temperature and 2.8 of mean molecular weight, the critical line mass should be 16 $M_{\odot }/\rm pc$.
We find $\sim $70\% of robust 
prestellar cores (135/199) embedded in supercritical filaments 
which implies that the gravitationally bound cores come from 
fragmentation of supercritical filaments.

\section{Conclusions}
\label{sect:conclusion}
With the latest improved difference term algorithm: hires, we made a 
high-resolution (${13.5}''$) column density map for Perseus MC with
$Herschel$ multi-wavelength dust continuum maps, and detected the source 
and filamentary structure using a new spatial decomposition method: $getsf$, 
and performed statistics on measured physical parameters so as to better understand 
the initial conditions  of the star formation in the molecular cloud. 
Our findings can be summarized as follows:
\begin{itemize}
\item  
We find power-law distribution in PDF of the Perseus south region 
is flatter than north region, and the average temperature of dense gas component with 
$N_{\rm H_{2}}>3\sigma$ in south region is about 5 K 
lower than the north region, 
and the number of YSOs in the south is significantly less than 
that in the north. Those observational evidences implies that 
south region is more gravitationally bound than north region. 
and suggests that evolution stages are different in two regions.
\item 
We selected 952 reliable cores from original source catalog detected by $getsf$, 
and divided them into four groups: 536 unbound starless cores, 87 candidate 
prestellar cores and 199 robust prestellar cores, 130 protostellar cores. 
We find $M$ strongly correlate with $T$ and $M/L$ strongly 
correlate with $M/T$, which is in line with 
the prediction of the blackbody radiation.
These two correlations shows a clear evolutionary sequence from unbound 
starless cores to robust prestellar cores.
\item 
We find crest $N_{\rm H_{2}}$ of the filamentary structures
are clearly related with $M_{\rm line}$, and are uncorrelated 
with $W$ by three orders of magnitude from $\sim 10^{20}$ to 
$\sim 10^{22}$ $\rm cm^{-2}$. 
We find $\sim $ 70\% of robust prestellar cores (135/199) embedded 
in supercritical filaments with $M_{\rm line} >16~M_{\odot}/{\rm pc}$,
which implies that the gravitationally bound cores come from fragmentation 
of supercritical filaments.
\end{itemize}
\begin{acknowledgements}
We would like to thank the anonymous referee for valuable comments which improved the quality of the paper.
This work is supported by the Ministry of Science and Technology of China through grant 2010DFA02710, the Key Project of Interntional 
Cooperation, and by the National Natural Science Foundation of China through grants 11503035, 11573036, 11373009, 11433008, 
11403040 and 11403041.
Guoyin ZHANG acknowledges support from China Postdoctoral Science Foundation (No. 2021T140672),
and National Natural Science foundation of China (No. U2031118). 
\end{acknowledgements}

\bibliography{ref}
\bibliographystyle{raa} 
\end{document}